\documentclass[a4paper,11pt]{article}
\usepackage{ragged2e}
\usepackage{floatrow}
\newfloatcommand{capbtabbox}{table}[][\FBwidth]

\usepackage{jinstpub} 
\usepackage[section]{placeins}
\usepackage[toc,page]{appendix}
\usepackage{graphicx}
\usepackage{subcaption}

\usepackage[english]{babel}

\usepackage{tikz}
\usetikzlibrary{patterns}
\usetikzlibrary{plotmarks}

\usepackage{amsmath}	
\usepackage{amssymb}
\usepackage{upgreek}
\usepackage{siunitx}

\usepackage{verbatim}
\usepackage{float}
\usepackage{calligra}
\usepackage{mathrsfs}
\usepackage{tabularx, booktabs}
\makeatletter
\newcommand*{\rom}[1]{\expandafter\@slowromancap\romannumeral #1@}
\makeatother
\usepackage{graphicx}
\usepackage{placeins}
\usepackage{wrapfig}
\usepackage{graphicx}
\usepackage{caption}
\usepackage{changepage}

\usepackage{float}
\floatstyle{plaintop}
\restylefloat{table}

\usepackage{hyperref}
\hypersetup{
    citecolor=red,
    linkcolor=blue,
    urlcolor=blue
}

\graphicspath{ {./images/} }
\title{Fast calculation of capacitances in silicon sensors with  3D and 2D numerical solutions of the Laplace's equation and comparison with experimental data and TCAD simulations}

\author[a,1]{P. Assiouras,\note{Corresponding author.}}
\author[a]{P. Asenov,}
\author[a]{A. Kyriakis}
\author[a]{and D. Loukas}

\affiliation[a]{Institute of Nuclear and Particle Physics (INPP), NCSR Demokritos \\Agia Paraskevi, Greece}

\emailAdd{panagiotis.assiouras@cern.ch}

\abstract{We have developed a software for fast calculation of capacitances in planar silicon pixel and strip sensors, based on 3D and 2D numerical solutions of the Laplace’s equation. The validity of  the 2D calculations was checked with  capacitances measurements on  Multi-Geometry Silicon Strip Detectors (MSSD). The 3D calculations were tested by comparison with pixel sensors capacitance measurements from literature. In both cases the Laplace equation results were compared with simulations obtained from the TCAD Sentaurus suite. The developed software is a useful tool for fast estimation of interstrip, interpixel and backplane capacitances, saving computation time as a first approximation before using  a more sophisticated platform for more accurate results if needed.
}


\begin{document}
	\maketitle
	\flushbottom
	
	\section{Introduction}
	\label{sec:intro}

Silicon sensors are extensively used in High Energy Physics experiments as tracking detectors of charged particles. The most commonly used planar silicon detectors in High Energy Physics experiments are devices segmented into strips (micro-strip detectors) or pixels (micro-pixel detectors). Important parameters in the operation and the design of silicon detectors are the capacitances between adjacent strips or pixels and between the strips or pixels and the backplane. These capacitances are related to signal to noise ratio  as well as crosstalk phenomena between neighboring pixels or strips.

A numerical algorithm for solving the three dimensional Laplace's
equation and calculating the capacitances of a pixel sensor, was presented in \cite{kavadias}. A reduced form of the  algorithm has been implemented for calculating the capacitances of micro-strip sensors by solving the Laplace's equation in two dimensions. Through these algorithms, numerical calculations of the capacitances between adjacent strips or pixels as well as the capacitances between the strips or pixels and the backplane can be made. These algorithms have been implemented within a software that can be used as a simulation tool for a fast estimation to lower order of the above mentioned capacitances.

Calculations for the pixel sensors obtained with our method are compared with published experimental data and TCAD simulations on various configurations of pixel geometries. Calculations for  strip sensors, are compared with experimental results and TCAD simulations on multi-geometry strip sensors (MSSD). The MSSD sensors were kindly provided by the Outer Tracker Sensor working group for the Phase-2 upgrade of the  CMS/LHC collaboration \cite{CMS3}.	 
 		  
\section{Numerical solution of the Laplace's equation}

 In the current section we give an outline of the method,  for a detailed presentation  see \cite{kavadias}. A key characteristic of this method is that the axes that are parallel to the pixel or strip plane are discretized in finite elements, while the perpendicular axis is kept continuous. Then by using a Fourier transform, the three dimensional problem for the pixel sensors is reduced to two dimensions and the two dimensional problem for the strip sensors is reduced to one dimension. The problem is then solved in Fourier space by using a numerical method.

\subsection{Three dimensional solution}
\label{sbsc:numerical_solution_pixel_sensors}

\begin{figure}[!hbtp]
    \centering
   \includegraphics[height=5cm, width=0.5\textwidth]{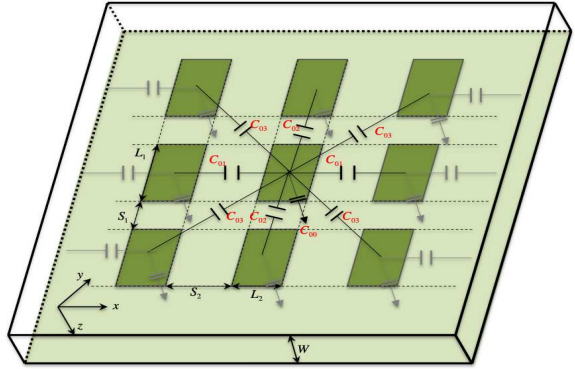}
   \caption[Schematic of the capacitive network]{Schematic of the capacitive network of a pixel sensor with 9 pixels. $C_{01}$ and $C_{02}$ are the capacitances that are formed between the central and the adjacent pixels in the directions that are parallel to the $ x$- and $ y$- axis respectively. $C_{03}$ are the capacitances that are formed between the central and the adjacent pixels in the diagonal direction and $C_{00}$ are the capacitances that are formed between each pixel and the backplane.}
   \label{fig:pixel9rectangle}
\end{figure}

Figure \ref{fig:pixel9rectangle} shows the capacitive network of a pixel detector. The capacitances that are calculated with this method are those formed between the central pixel and the adjacent pixels in the directions that are parallel to $x$- and $y$- axis respectively ($C_{01}$ and $C_{02}$), the capacitances that are formed between the central pixel and the adjacent pixels in the diagonal direction ($C_{03}$) and the capacitances that are formed between each pixel and the backplane ($C_{00}$). These capacitances are strongly related to the geometry features of the sensor such as the dimensions of each pixel and the separation gap between them and the thickness.

To calculate the strip sensor capacitances the Poisson's equation is solved with normalized boundary conditions by setting $V(x,y,z)=1$ at the central pixel and $V(x,y,z)=0$ at the adjacent pixels and the backplane. In the areas not covered by pixel the equation ($\varepsilon_{Si} E_{Si}(x,y,z)-\varepsilon_{a} E_{a}(x,y,z)=0$) is applied for keeping the electric field in the interface continuous, where $\varepsilon_{Si}$, $\varepsilon_{a}$ are the dielectric constants and $E_{Si}(x,y,z)$, $E_{a}(x,y,z)$ are the vertical components of the electric field in silicon and ambient space, respectively. The ambient space in this work is considered to be air. 

The detector is considered to be in a fully depleted state and it is free from thermally generated free charge carriers (pairs of electrons and holes). This is the state in which  a realistic silicon detector works, when a reverse bias voltage is applied to the silicon sensor and the depleted region is formed. The thermally generated free charged pairs of electrons and holes are swept from the electric field creating an ionization chamber. The pixels are  assumed to be infinitesimally small in depth compared to the fully depleted region of the detector. Also, for simplicity it is considered that the  volume and the surface of the detector are free from static charges. Under real circumstances, charges exist inside the detector volume. These are stripped ions in the depleted regions and defects from contamination inside the material. However, they produce an electric field component which is independent of the
biasing voltage. The charges add a voltage-independent term in the expression of the strip charge
which does not influence the calculation of the capacitance.

Under the aforementioned assumptions  Poisson's equation is reduced to the Laplace's equation \ref{eq:laplace_equation}. Subsequently, by using Fourier transform, while keeping the perpendicular axis continuous,  the three dimensional problem is reduced to two dimensions (equation \ref{eq:diffential_equation}). For details see ref. \cite{kavadias}. 

\begin{align}
\label{eq:laplace_equation}
\nabla^{2}V(x,y,z) &=0  \xrightarrow{\mathcal{F}} 
\\ 
\frac{\partial^2 {V(k_{x},k_{y},z)}}{\partial z^2} &=\left( k_{x}^{2}+k_{y}^{2}\right) V\left( k_{x},k_{y},z\right)
\label{eq:diffential_equation}
\end{align}

where $V\left( k_{x},k_{y},z\right)$  is the potential in Fourier space and $k_{x}$, $k_{y}$ the corresponding coordinates in Fourier space. By solving the differential equation \label{eq:diffential_laplace_equation_fourier} 
with the appropriate boundary conditions,  equation \ref{eq:function_of_electric_field_with_potential} is derived which gives the electric field in Fourier space ($E_{z}\left( k_{x},k_{y},0\right)$) as a function of the potential in Fourier space ($V\left( k_{x},k_{y},0\right)$), which for $z=0$  corresponds to the pixel plane, where w corresponds to the detector thickness and $F\left( k_{x},k_{y}\right)$ is a function of the Fourier coefficients.

\begin{equation}\label{eq:function_of_electric_field_with_potential} 
E_{z}\left( k_{x},k_{y},0\right)=F\left( k_{x},k_{y}\right) \
\frac{1+e^{-2F\left( k_{x},k_{y}\right)w}}{1-e^{-2F\left( k_{x},k_{y}\right)w}} V\left( k_{x},k_{y},0\right)  
\end{equation}

The Laplace's equation is then solved in the Fourier space by using a self-consistent numerical method. First an initial guess of the potential is made $V_{init}(x,y,0)$ where for $(z=0)$ it corresponds to the pixel plane. By using Fourier transform, with respect to $x$- and $y$- axis the initial guess is transformed to the potential in Fourier space $V_{init}(k_{x},k_{y},0)$. Then the vertical component of the electric field inside the sensor $E^{init}_{Si}(x,y,0)$ is calculated from \ref{eq:function_of_electric_field_with_potential}  by performing inverse Fourier transform. The  vertical component of the electric field in the ambient space $E^{init}_{a}(x,y,0)$ is calculated by equation \ref{eq:function_of_electric_field_with_potential} setting ($w \rightarrow \infty$) and using inverse Fourier transform. Next the potential is redefined  by boundary conditions. The values of the vertical component of the electric field in the ambient space $E^{init}_{a}(x,y,0)$ are used for calculating the new values of the electric field inside the sensor $E^{init}_{Si}(x,y,0)$, by using the boundary condition in the space that is not covered by pixels. This gives a new estimation of the electric field inside the sensor $E_{new}(x,y,0)$ and a new estimation of the potential $V_{new}(x,y,0)$. The actual solution of the problem is assumed to be a linear combination of the new and the initial potential functions. Then a check for convergence is made and if it has not been achieved the initial function is set equal to the new potential functions, that has been derived from the linear combination. All the above steps are repeated through several cycles until convergence is reached. 

Finally, once convergence has been achieved, the charges stored in each pixel are calculated by integrating  the charge density in the whole pixel surface ($Q= \oiint (\varepsilon_{Si}E_{Si}-\varepsilon_{a}E_{a} )dS$). Then the calculation of pixel capacitances is made.

\subsection{Two dimensional solution}
\label{sbsc:numerical_solution_strip_sensors}

\begin{figure}[!htbp]
    \centering
   \includegraphics[height=4 cm]{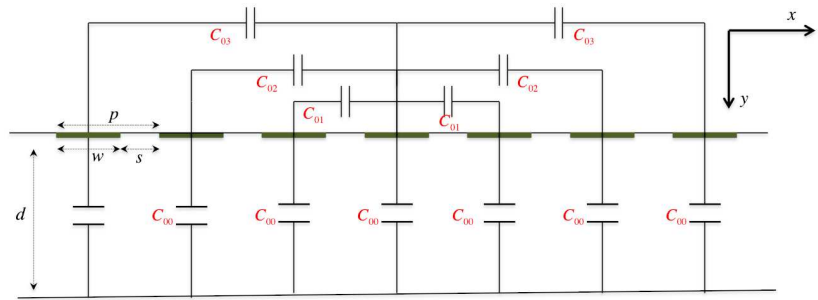}
   \caption[]{ Schematic of the capacitive network of a micro-strip detector with 7 strips. $C_{01}$ are the capacitances between the central and the first adjacent strip, $C_{02}$ between the central and the second adjacent strip, $C_{03}$ between the central strip and the third adjacent strip, and $C_{00}$ between each strip and the backplane, where w is the strip width, s is the interstrip space and p is the strip pitch and d is the detector thickness}
   \label{fig:5stripdector}
\end{figure}

This algorithm can be used for calculating the capacitances in micro-strip detectors with planar geometry. Figure \ref{fig:5stripdector} shows the capacitive network of a strip sensor with 7 strips.

The capacitances that are calculated with this algorithm are those between a central strip and the backplane ($C_{00}$), the capacitances between the central and the first adjacent strips ($C_{01}$) and between the central and the second adjacent strips ($C_{02}$). While the software calculates capacitances up to the third  ($C_{03}$), these are negligible and omitted from our study. The same method as in the three dimensional case is followed, with the difference that in the strip sensor case the problem is reduced to two dimensions and by using Fourier transform the Laplace's equation is solved in one dimension. In this case the axis that is parallel to the strip plane ($x$- axis) is discretized in finite elements, while the perpendicular ($y$- axis) is kept continuous.

\section{TCAD Simulations}
		
Technology computer-aided design (TCAD) is used in the semiconductor industry in order to develop and optimize semiconductor processing technologies and devices. It can be used in order to simulate the fabrication procedure, the operation and the reliability of the semiconductor devices. The TCAD suite that was used for this work is the commercial software package TCAD Sentaurus from Synopsys \cite{TCAD}.
		
TCAD follows a finite element analysis scheme. Firstly, the device is designed in two or three dimensions and the properties of each region of the device, such as the doping concentration, the materials or shape are defined. Another way to create a device is by simulating the actual fabrication procedure,  but this approach is beyond the scope of the present work. Afterwards the device is subdivided into finite elements by following a Delaunay triangulation algorithm \cite{delaunay} which creates a mesh of the device.

The next step is to activate the desired physical models and parameters before initiating the device simulation program. Some of the physical models that were used in this work are the Auger recombination, Shockley-Read-Hall recombination, avalanche electron-hole generation, trap-to-trap recombination, band-to-band tunneling, doping dependence mobility, high field saturation and carrier-carrier scattering \cite{TCADmanual}.  

The fundamental partial differential equations for semiconductors (Poisson's, continuity equations for electrons and holes) are solved  at each of the generated mesh point and the desired physical quantities are calculated. In order to calculate the capacitances a small signal AC analysis is performed at 1 kHz for the backplane capacitance and at 1 MHz for the interstrip capacitance. These frequencies correspond to the frequencies that the experimental measurements of this sensors were performed.

\subsection{Strip sensors}
\label{sbsc:TCAD_simulations_strip_sensors}

\begin{figure}[!htb]
	\centering
	\includegraphics[height=10.0cm,width=0.9\textwidth]{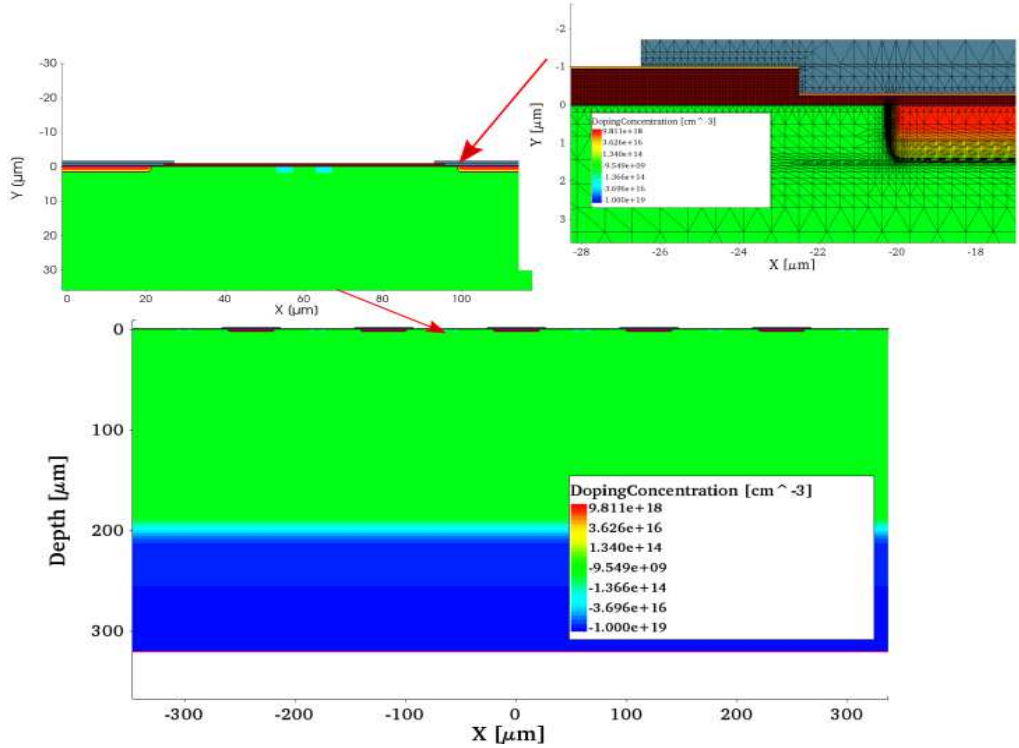}
	\caption[]{
		The structure that was used for simulating the MSSD sensors. In the bottom middle the whole structure that was used with 5 strips is shown. The color variation depicts the doping concentration. The top-left figure shows a close up view between two strips. Two p-stop structures are designed, depicted in cyan, in order to achieve strip isolation. The top-right figure shows a close-up view of the simulated structure near to one strip edge. The n-type implant is
		displayed in red which is simulated with a Gaussian profile, the aluminum contacts are displayed
		with gray, the $SiO_{2}$ is displayed with brown and the p-type silicon bulk is displayed with green. }
	\label{fig:mssd_TCAD}
\end{figure}

Figure \ref{fig:mssd_TCAD} shows the simulated structure of the MSSDs. The design resembles a perpendicular cross-section of the sensor to the strip plane. The final results are scaled to the actual sensor strip length. The structure has 5 strips instead of 32 of the actual MSSD sensors with implant and aluminum widths as denoted in table \ref{tb:MSSD_table}. 
The different layers of each strips are depicted in the top-right part of figure \ref{fig:mssd_TCAD} as well as the generated mesh consisting of triangular segments. The metal depicted in gray is extended a few \si{\micro\metre} in the interstrip space. This technique is called metal-overhang and is used in order to overcome the junction curvature effect which limits the breakdown voltage of planar junctions \cite{SZE}, \cite{Passeri}. In the space between two strips two additional structures (depicted in cyan) have been designed with high dose of p-implant resembling the actual p-stop structures of the sensor. The bulk doping concentration is assumed to be equal to $3.5 \times 10^{12}$ \si{\centi\metre^{-3}} (p-type), while the strip doping concentration is assumed to be  equal to $1.0 \times 10^{19}$ \si{\centi\metre^{-3}} (n-type). The deep diffusion technique on the backplane is simulated by using an error function doping profile. More details on the parameters that were used in order to produce these simulations can be seen in Appendix \ref{Appendix A}. Most of the parameters that are used in the simulation have  been chosen by following the works presented in \cite{Eichhorn:2015vea} and \cite{Eber2013_1000038403}.

The interstrip capacitance between two strips i and j for an AC coupled sensor  are calculated, according to \cite{CHATTERJI20031491}, with the following formula \ref{eq:Cint}.
\begin{equation}
C_{int}=
C_{M_{i}-M_{j}} + C_{I_{i}-I_{j}} + C_{M_{i}-I_{j}} + C_{I_{i}-M_{j}}
\label{eq:Cint}
\end{equation}

where $C_{M_{i}-M_{j}}$ is the capacitance between the metal of the  i\textsuperscript{th} strip and the metal of the j\textsuperscript{th} strip, $C_{I_{i}-I_{j}}$ is the capacitance between the implant of i\textsuperscript{th} strip and the implant of j\textsuperscript{th} strip, $C_{M_{i}-I_{j}}$ is the capacitance between the metal of i\textsuperscript{th} strip and the implant of j\textsuperscript{th} strip, $C_{I_{i}-M_{j}}$ is the capacitance between the implant of i\textsuperscript{th} strip and the metal of j\textsuperscript{th} strip.

\subsection{Pixel sensors}
\label{sbsc:TCAD_pixel_sensors}
\begin{figure}[htbp!]
	\centering
	\begin{subfigure}[htbp!]{0.45\textwidth}
		\centering
		\includegraphics[height=3.0cm,width=\textwidth]{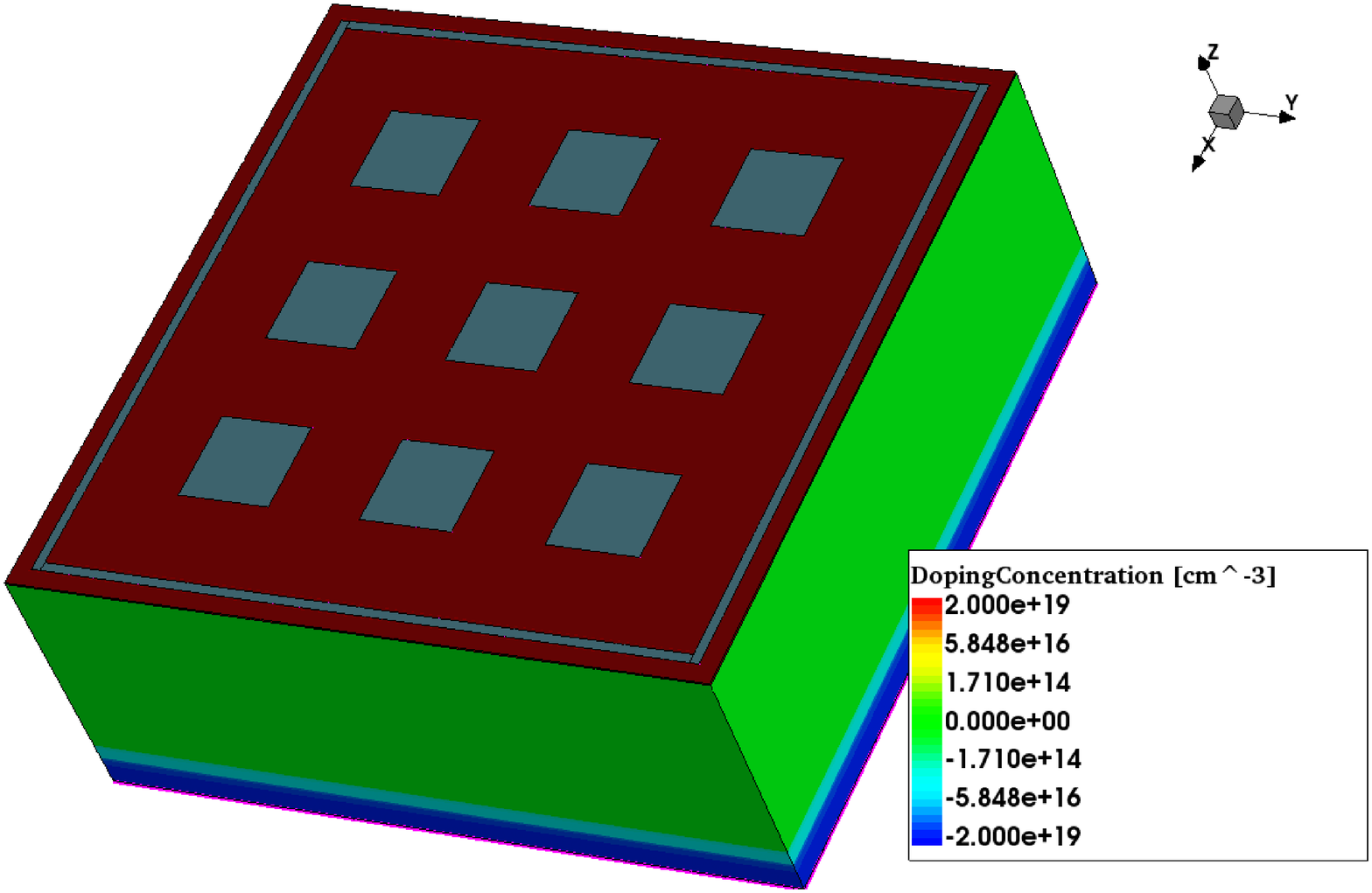}
		\caption{\label{fig:pixelSensor_50x50} }  
	\end{subfigure}
	\quad
	\begin{subfigure}[htbp!]{0.40\textwidth}
		\centering
		\includegraphics[height=3.0cm, width=\textwidth]{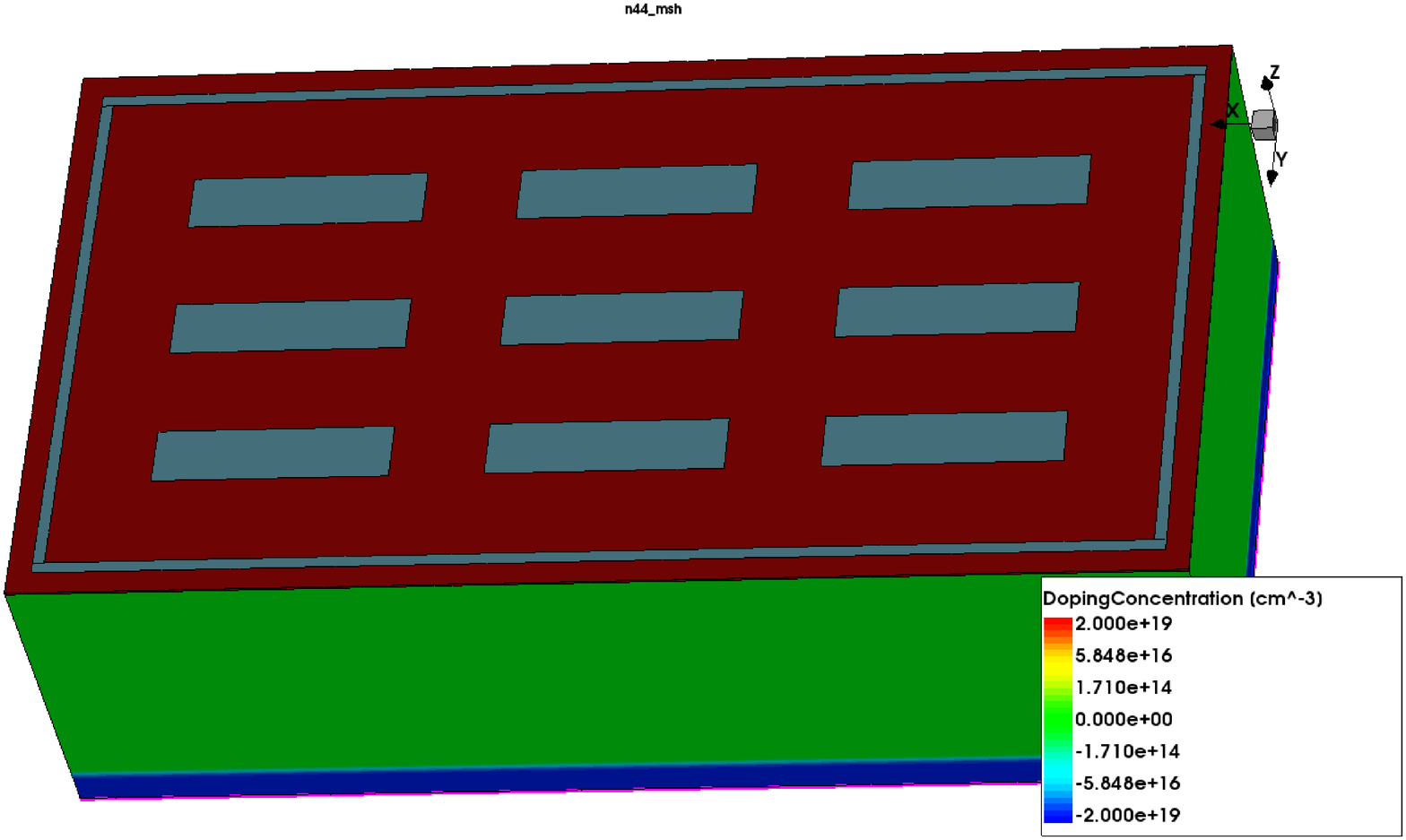}
		\caption{\label{fig:pixelSensor_100x25}}
	\end{subfigure}
	\caption{Simulated structure for pixel sensors with $50\times50$ \si{\micro\meter^2}, (figure \ref{fig:pixelSensor_50x50}) and with $100\times25$ \si{\micro\meter^2}, (figure \ref{fig:pixelSensor_100x25}) pixel area.}
\end{figure}

For pixel sensors a 3D simulation approach is employed. The simulated structures  consist of 9 orthogonal pixels with two different pixel geometries one with pixel area of  $50\times50$  \si{\micro\meter^2}, (figure \ref{fig:pixelSensor_50x50}), and one with pixel area of $100\times25$ \si{\micro\meter^2}, (figure \ref{fig:pixelSensor_100x25}). Both structures are DC-coupled with a n\textsuperscript{+}p configuration and an active thickness of $150$ \si{\micro\meter}. A guard ring structure surrounds the device, providing a homogeneous electric field inside the sensitive area and minimizing the edge effects. The simulations were made for different pixel layouts with varying separation gap between $5$ \si{\micro\meter} to $50$ \si{\micro\meter} with a $5$ \si{\micro\meter} step. The capacitances were calculated by performing a small signal AC analysis with frequencies at 1 kHz and 1 MHz for the backplane and interpixel capacitances, respectively. These are the configurations under development for the Phase-2 upgrade of the CMS/LHC \cite{CMS3} and Atlas/LHC \cite{Atlas-Pixel} silicon trackers at CERN. Simulation parameters are shown in Appendix \ref{Appendix A}.

\section{Experimental measurements and comparison with simulation}
\label{sec:Experimental_mesauments}
\subsection{Geometry of strip sensors and experimental setup}

\begin{figure}[!htbp]
	\centering
	\includegraphics[height=3.5cm, width=\textwidth]{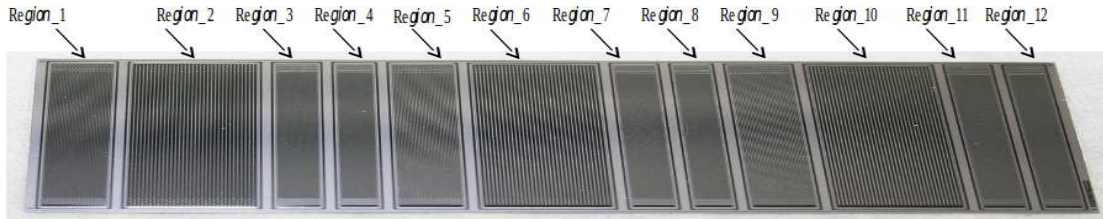}
	\caption[Multi-Geometry Silicon Strip Sensor]{A Multi-Geometry Silicon Strip Sensor. The sensor has 12 regions with different pitches and width-to-pitch ratio (w/p). Three of these sensors with physical thickness of 320 \si{\micro\meter} and active thickness of 120, 200 and 320 \si{\micro\meter} used for the measurements}
	\label{fig: MSSD}
\end{figure}

Figure \ref{fig: MSSD} shows an actual picture of a Multi Geometry Silicon Strip Detector (MSSD). The MSSDs contain $12$ individual regions and they all have their own bias and guard rings. Each region contains $32$ AC coupled strips on n\textsuperscript{+}p configuration with pitches varying from $70$ to $240$ \si{\micro\meter} resulting to width-to-pitch ratios (w/p) varying from $0.133$ to $0.321$. The geometrical characteristics of each region are described in table \ref{tb:MSSD_table}. More details for these sensors can be found in \cite{HOFFMANN201130}. The multi-geometry of these sensors allows the study of capacitances with different width to pitch ratios to be made. This makes them a suitable choice for checking the calculation capability of our software for various strip geometries. 

\begin{table}[!htpb]
	\begin{flushleft}
		\resizebox{1.0\textwidth}{!}{%
			\begin{tabular}{|| c || c | c | c | c || c | c | c | c || c | c | c | c ||} 
				\hline 
				Label & $1(120)$ & $1(240)$ & $1(80)$ & $1(70)$ & $2(120)$ & $2(240)$ & $2(80)$ & $2(70)$ & $3(120)$ & $3(240)$ & $3(80)$ & $3(70)$ \\ [0.5ex] 
				\hline \hline
				Region & $1$ & $2$ & $3$ & $4$ & $5$ & $6$ & $7$ & $8$ & $9$ & $10$ & $11$ & $12$ \\ 
				\hline
				Pitch & 120 & 240 & 80 & 70 & 120 & 240 & 80 & 70 & 120 & 240 & 80 & 70 \\
				\hline
				Width & $16$ & $34$ & $10$ & $8.5$ & $28$ & $58$ & $18$ & $15.5$ & $40$ & $82$ & $26$ & $22.5$ \\
				\hline
				Width of Al & $29$ & $47$ & $23$ & $21.5$ & $41$ & $71$ & $31$ & $28.5$ & $53$ & $95$ & $39$ & $33.5$  \\
				\hline
				w/p & $0.133$ & $0.142$ & $0.125$ & $0.121$ & 
				$0.233$ & 
				$0.242$ & $0.225$ & 
				$0.221$ & 
				$0.333$ & $0.342$ & $0.325$ & $0.321$  \\
				& $(0.142)$ & 
				$(0.146)$ & $(0.138)$ & $(0.136)$ & 
				$(0.242)$ & 
				$(0.246)$ & $(0.238)$ & 
				$(0.236)$ & 
				$(0.342)$ & $(0.346)$ & $(0.338)$ & $(0.336)$  \\[1ex] 
				\hline
				
			\end{tabular}
		}
	\end{flushleft}
	\caption[Multi-Geometry Silicon Strip Sensor]{Geometrical characteristics of the MSSD sensors for each region.
	}
	\label{tb:MSSD_table}
\end{table}
 The three MSSD sensors that we measured in this work originate from 3 different wafers of float zone silicon (FZ). They have the same physical thickness of 320 \si{\micro\meter},  but three different active thicknesses of 120 \si{\micro\meter} (FZ120P), 200 \si{\micro\meter} (FZ200P), 320 \si{\micro\meter} (FZ320P). The manufacturer achieves this by a deep diffusion technique. In order to achieve isolation of the strip implants an additional high dose of \textrm{p+} implantation is made between the strips surrounding each \textrm{n+} strip. These structures are called p-stop. 
 
The measurements were performed with  a semi-automated probe station (Carl Susse PA 150). The whole setup was electrically shielded inside a light-tight metal box. The capacitances are measured with an HP4192A LCR meter which supplies a small AC signal superimposed upon the DC bias voltage on the HIGH terminal. The amplitude and phase are measured on the LOW terminal. Backplane capacitances were measured  by using the bias ring which connects together all the 32 strips via the bias resistors. The measurements were performed at a frequency of 1 kHz with an amplitude of 250 mV. The bias voltage (detector HI) was applied to the backplane while the bias ring was grounded. The bias voltage was ramped up from 0 V to 400 V. The interstrip capacitances were measured by performing an automatic strip scan in each strip, of each region, of the sensor. In this measurement,two neighboring strips were contacted with the probes connected one with the HI terminal along with the backplane and the strip under test to the LOW terminal. The interstrip capacitance  measurements were performed at 1 MHz with an amplitude of 250 mV. The frequencies and the amplitudes that have been chosen correspond to those that yield the optimum C-V characteristics of these sensors with the particular experimental setup for the backplane and interstrip capacitances respectively.

\subsubsection{Strip backplane capacitances}

\begin{figure}[!htbp]
	\centering
	\begin{subfigure}{0.315\textwidth}
		\centering
		\includegraphics[height=3.5cm,width=\textwidth]{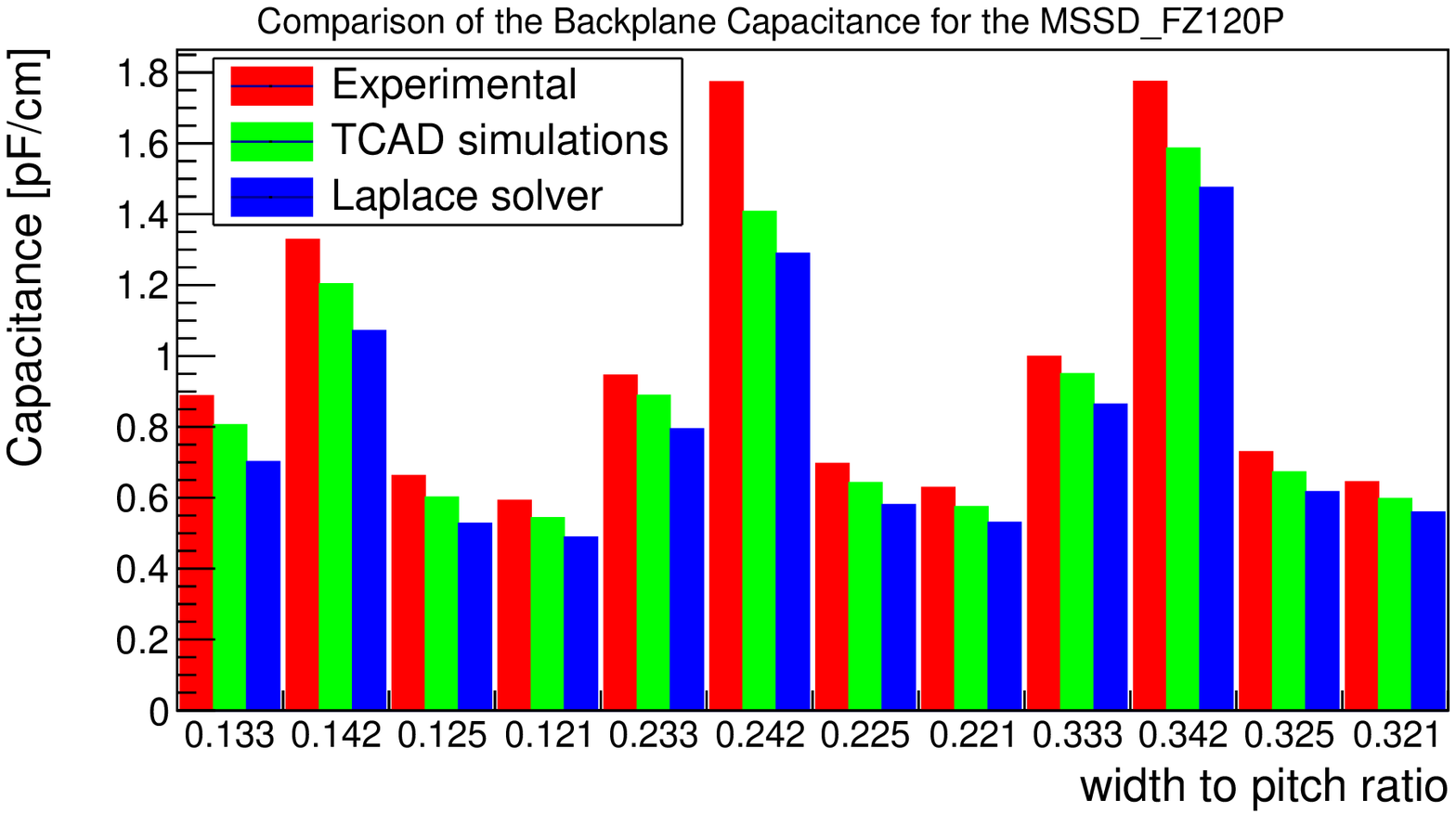}
		\caption{
			\label{fig:Comparison_Cback_MSSD_120P}}      
	\end{subfigure}\quad
	\begin{subfigure}{0.315\textwidth}
		\centering
		\includegraphics[height=3.5cm,width=\textwidth]{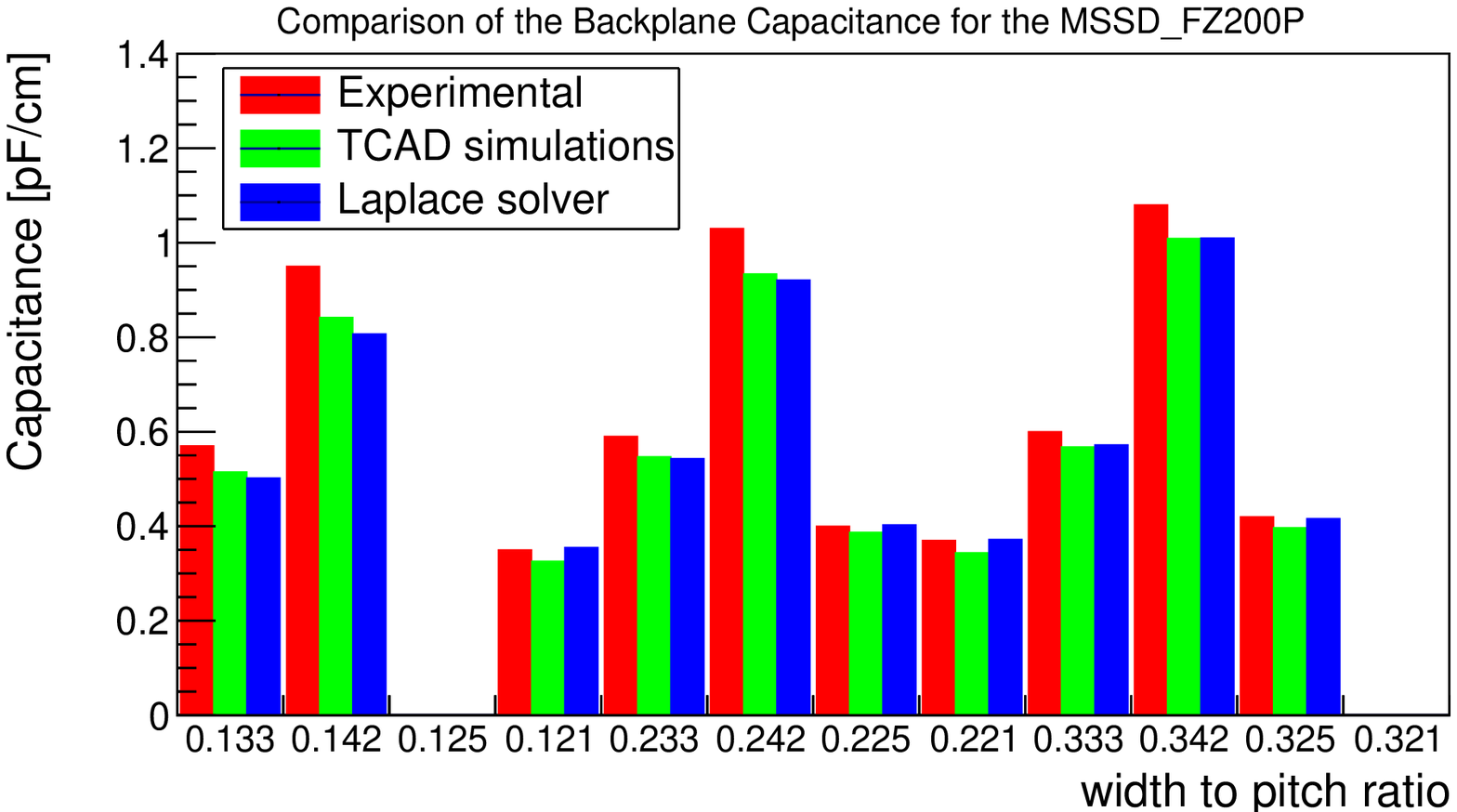}
		\caption{ \label{fig:Comparison_Cback_MSSD_200P}}
	\end{subfigure}\quad	\begin{subfigure}{0.315\textwidth}
		\centering
		\includegraphics[height=3.5cm,width=\textwidth]{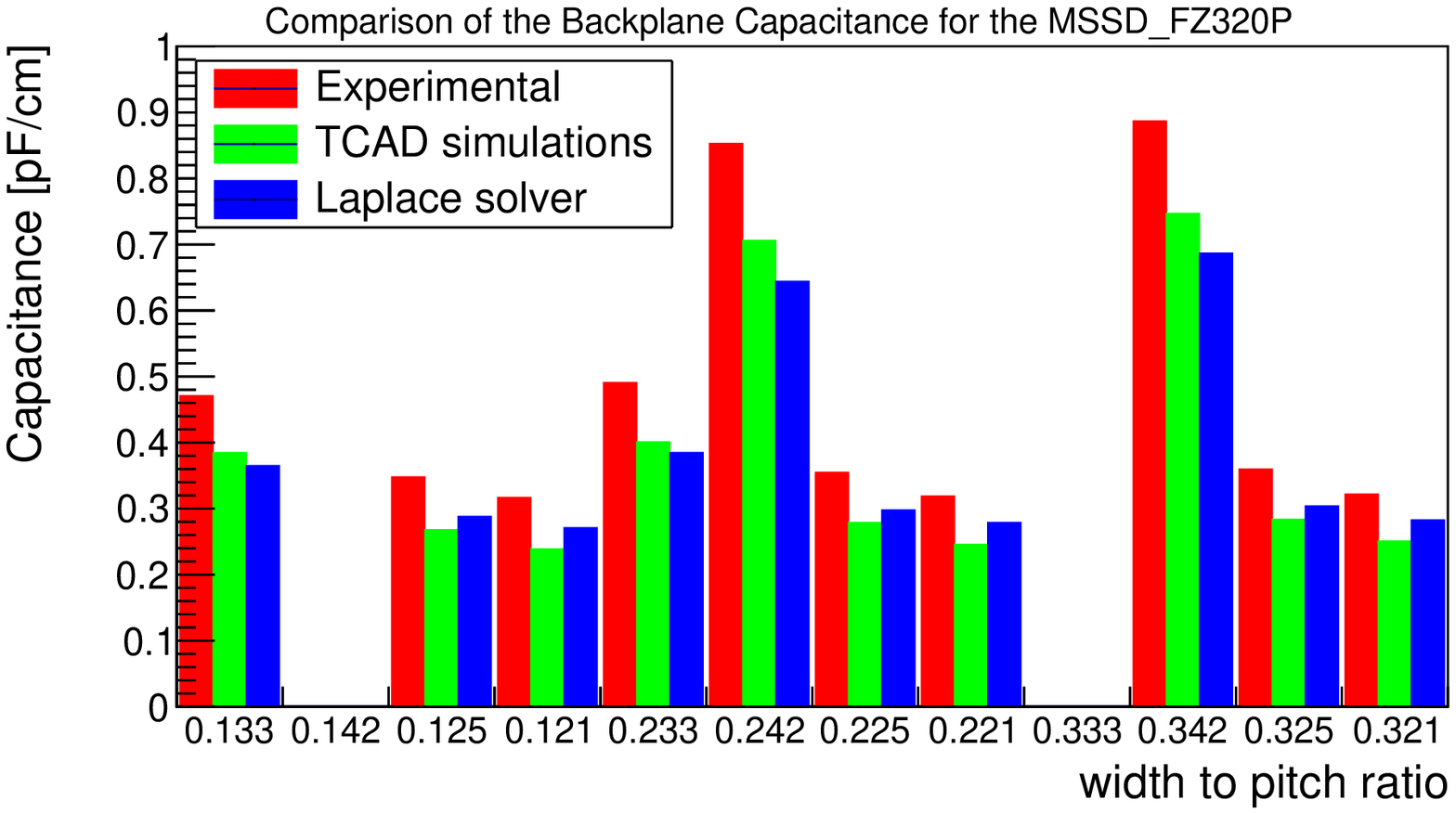}
		\caption{\label{fig:Comparison_Cback_MSSD_320P} }  
	\end{subfigure}
	\caption[]{Comparison of the experimental results (red), the results of the related TCAD simulations (green) and the Laplace solver (blue)  for the backplane capacitance in FZ120P \ref{fig:Comparison_Cback_MSSD_120P}, FZ200P \ref{fig:Comparison_Cback_MSSD_200P} and FZ320P
		\ref{fig:Comparison_Cback_MSSD_320P} sensors.
	}	
\end{figure}

Histograms \ref{fig:Comparison_Cback_MSSD_120P} , \ref{fig:Comparison_Cback_MSSD_200P} and \ref{fig:Comparison_Cback_MSSD_320P} show the  comparison between the backplane capacitances for the 3 MSSD sensors. The  regions $3$ and $12$ in the FZ200P sensor and $2$ and $9$ in the FZ320P sensor were damaged. In the majority of the measured samples with different w/p ratio the measured backplane capacitance is larger than the simulated one, probably due to parallel parasitic capacitances introduced during the measurement. The TCAD simulated capacitance seems to be closer to the measured one in the majority of our samples. However, our simpler but much faster Laplace solver gives quite comparable results for the backplane capacitance.

\begin{figure}[!htbp]
	\centering
	\begin{subfigure}{0.44\textwidth}
		\centering
		\includegraphics[height=4.5cm,width=\textwidth]{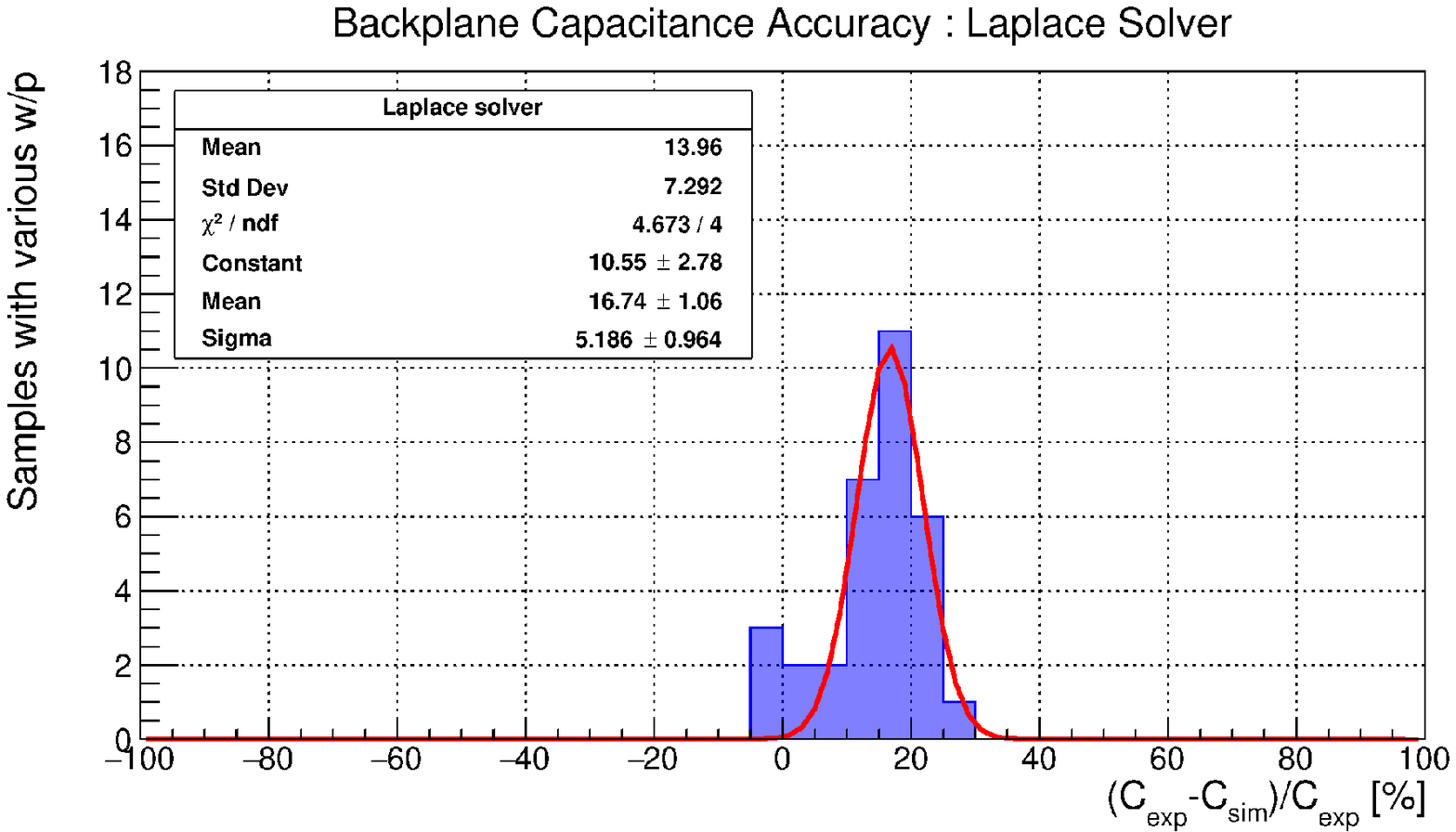}
		\caption{
			\label{fig:accuracy_histo_Cback_Laplce}}      
	\end{subfigure}\quad
	\begin{subfigure}{0.44\textwidth}
		\centering
		\includegraphics[height=4.5cm,width=\textwidth]{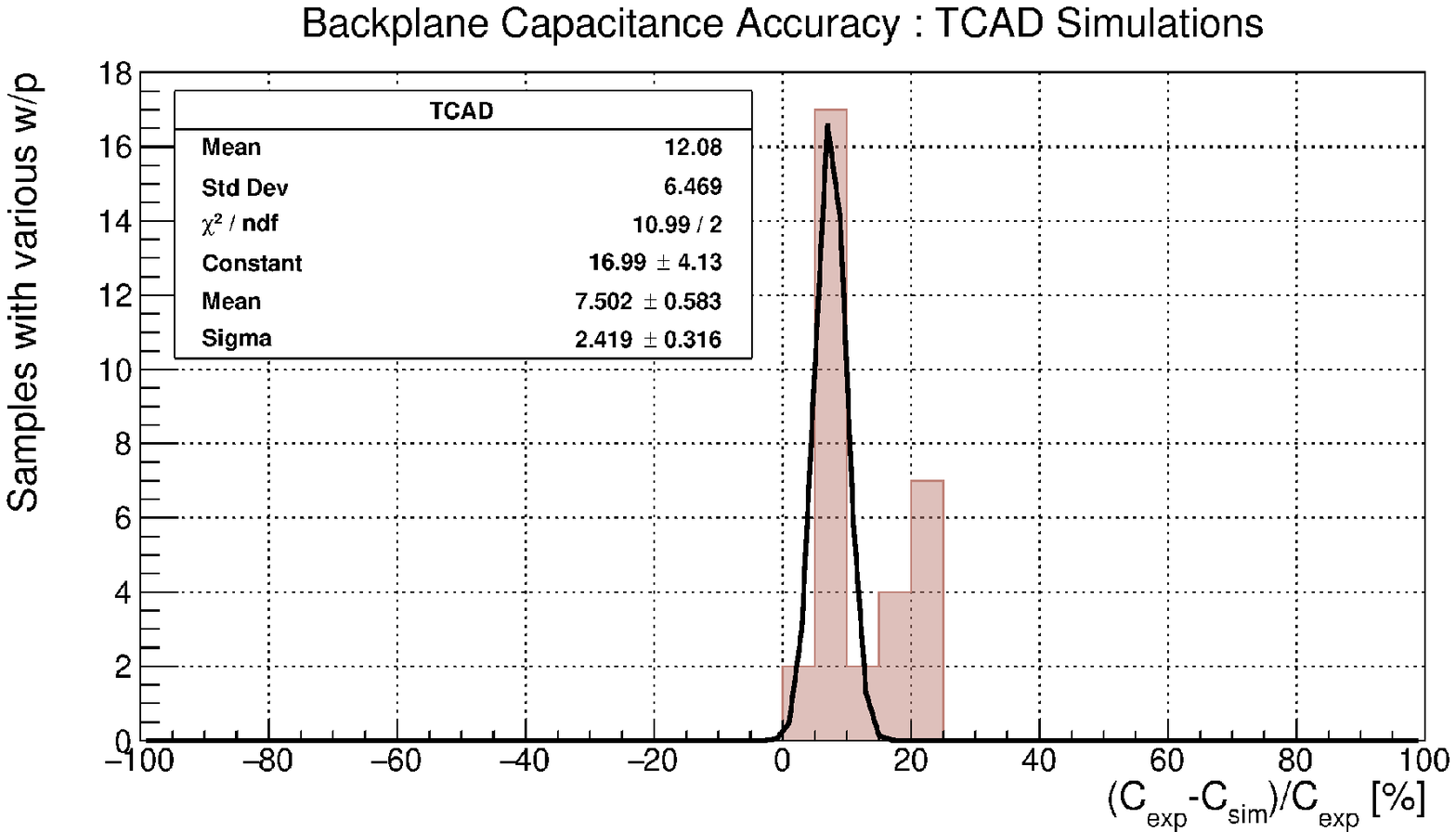}
		\caption{ \label{fig:accuracy_histo_Cback_TCAD}}
	\end{subfigure}
	\caption[]{Histograms of the relative error of the simulated results to the experimental values for the backplane capacitance for the Laplace solver \ref{fig:accuracy_histo_Cback_Laplce} and for the TCAD simulations \ref{fig:accuracy_histo_Cback_TCAD} from all 3 MSSD sensors}
\end{figure}

Figures \ref{fig:accuracy_histo_Cback_Laplce} and \ref{fig:accuracy_histo_Cback_TCAD} show the relative errors  of the backplane capacitance defined as $(C_{exp}-C_{sim})/C_{exp} \%$, of the backplane capacitance where $C_{sim}$ are the simulated results from our Laplace solver and the TCAD simulations respectively and $C_{exp}$ are the experimental values. A Gaussian fit is also shown for comparison. The numerical calculations made with the Laplace solver have a mean value of $14\%$ with a standard deviation of $7\%$ while the calculations made with TCAD have a mean value of $12\%$ and a standard deviation of $6\%$.

\subsubsection{Interstrip capacitances}

\begin{figure}[!htb]
	\centering
	\begin{subfigure}{0.31\textwidth}
		\centering
		\includegraphics[height=3.5cm,width=\textwidth]{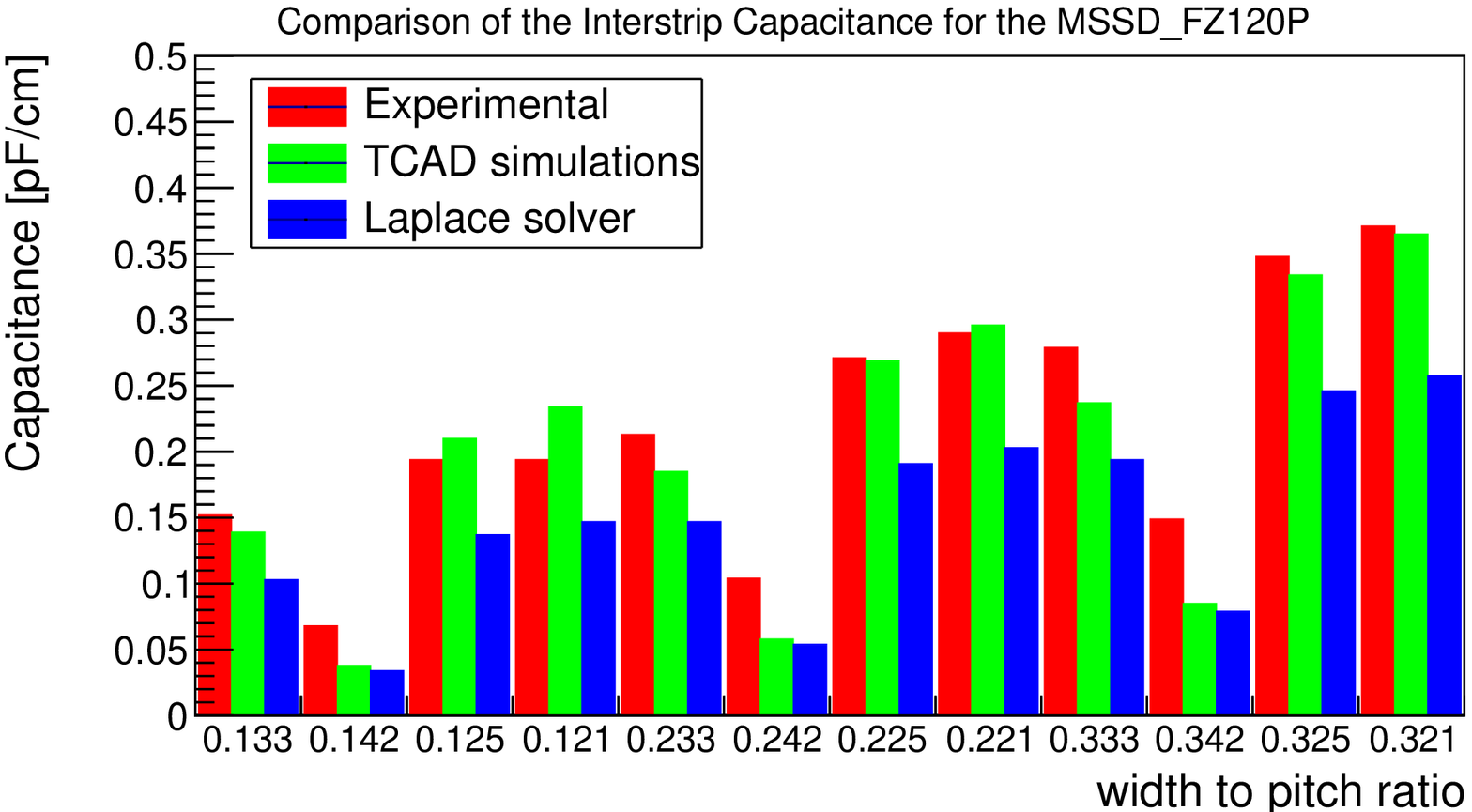}
		\caption{
			\label{fig:Comparison_Cint_MSSD_120P}}      
	\end{subfigure}
	\quad
	\begin{subfigure}{0.31\textwidth}
		\centering
		\includegraphics[height=3.5cm  , width=\textwidth]{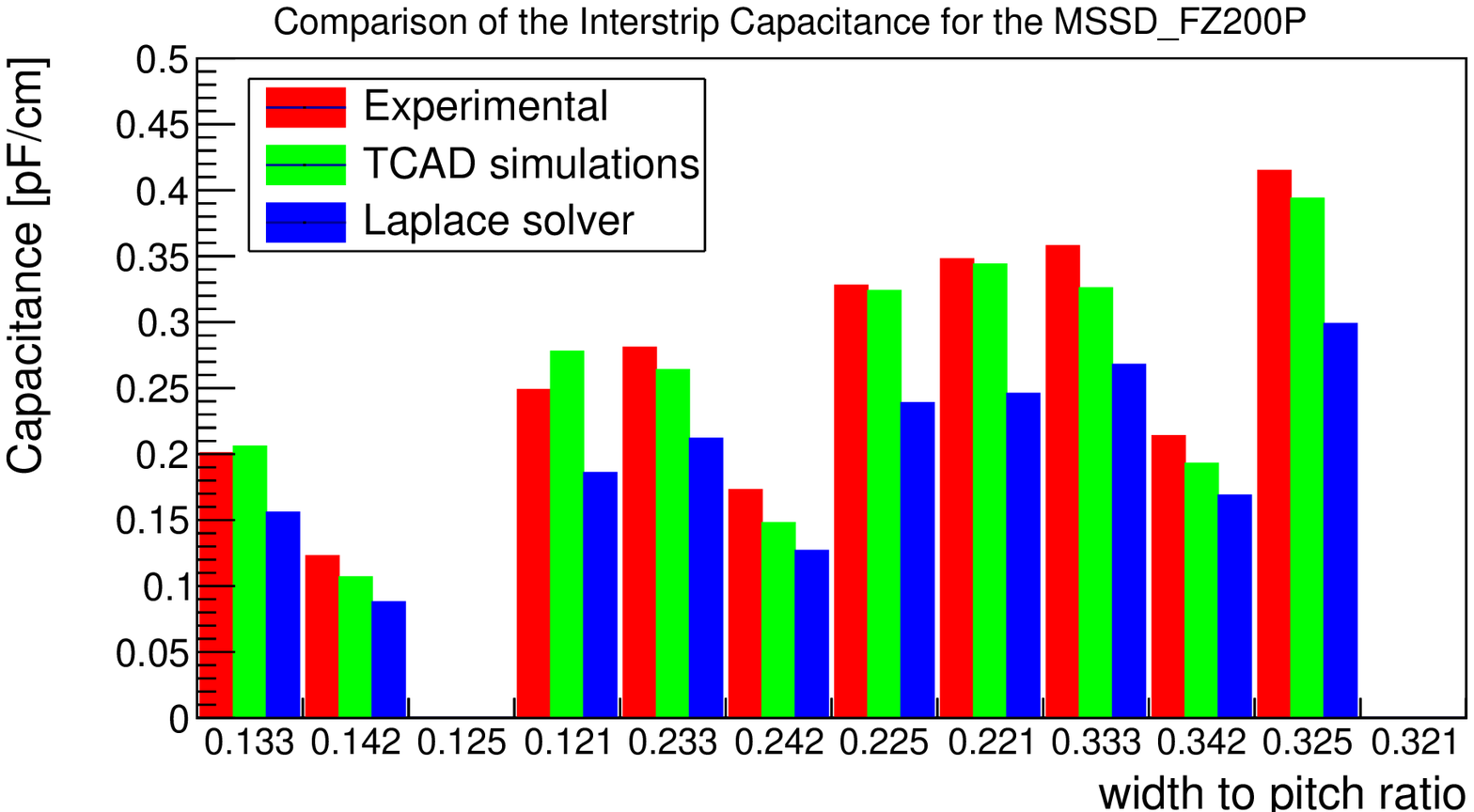}
		\caption{ \label{fig:Comparison_Cint_MSSD_200P}}
	\end{subfigure}
	\quad
	\begin{subfigure}{0.31\textwidth}
		\centering
		\includegraphics[height=3.5cm  , width=\textwidth]{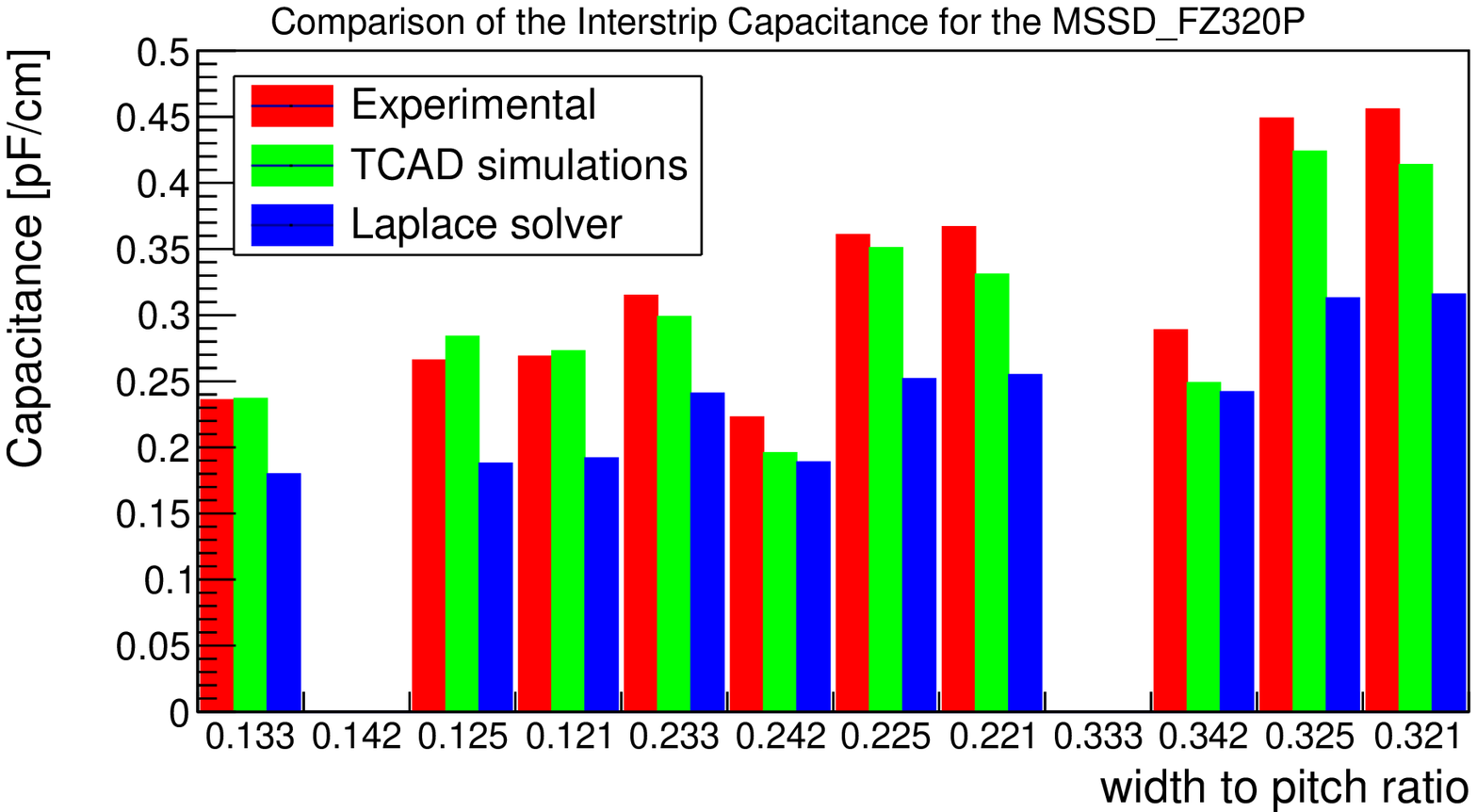}
		\caption{\label{fig:Comparison_Cint_MSSD_320P} }  
	\end{subfigure}
	\caption[]{Comparison of the experimental results  (red) the results of the related TCAD simulations (green) and the Laplace solver (blue) for the interstrip capacitance in FZ120P
		\ref{fig:Comparison_Cint_MSSD_120P}, FZ200P \ref{fig:Comparison_Cint_MSSD_200P} and FZ320P
		\ref{fig:Comparison_Cint_MSSD_320P} sensors.	}	
\end{figure}

Histograms \ref{fig:Comparison_Cint_MSSD_120P} , \ref{fig:Comparison_Cint_MSSD_200P} and \ref{fig:Comparison_Cint_MSSD_320P} show measured and simulated interstrip capacitances for the 3 MSSD sensors. As in the backplane case, in the majority of the measured samples with different w/p ratio the measured interstrip capacitance is larger than the simulated one, probably due to parallel parasitic capacitances introduced during the measurement.

\begin{figure}[!htbp]
	\centering
	\begin{subfigure}{0.44\textwidth}
		\centering
		\includegraphics[height=4.0cm,width=\textwidth]{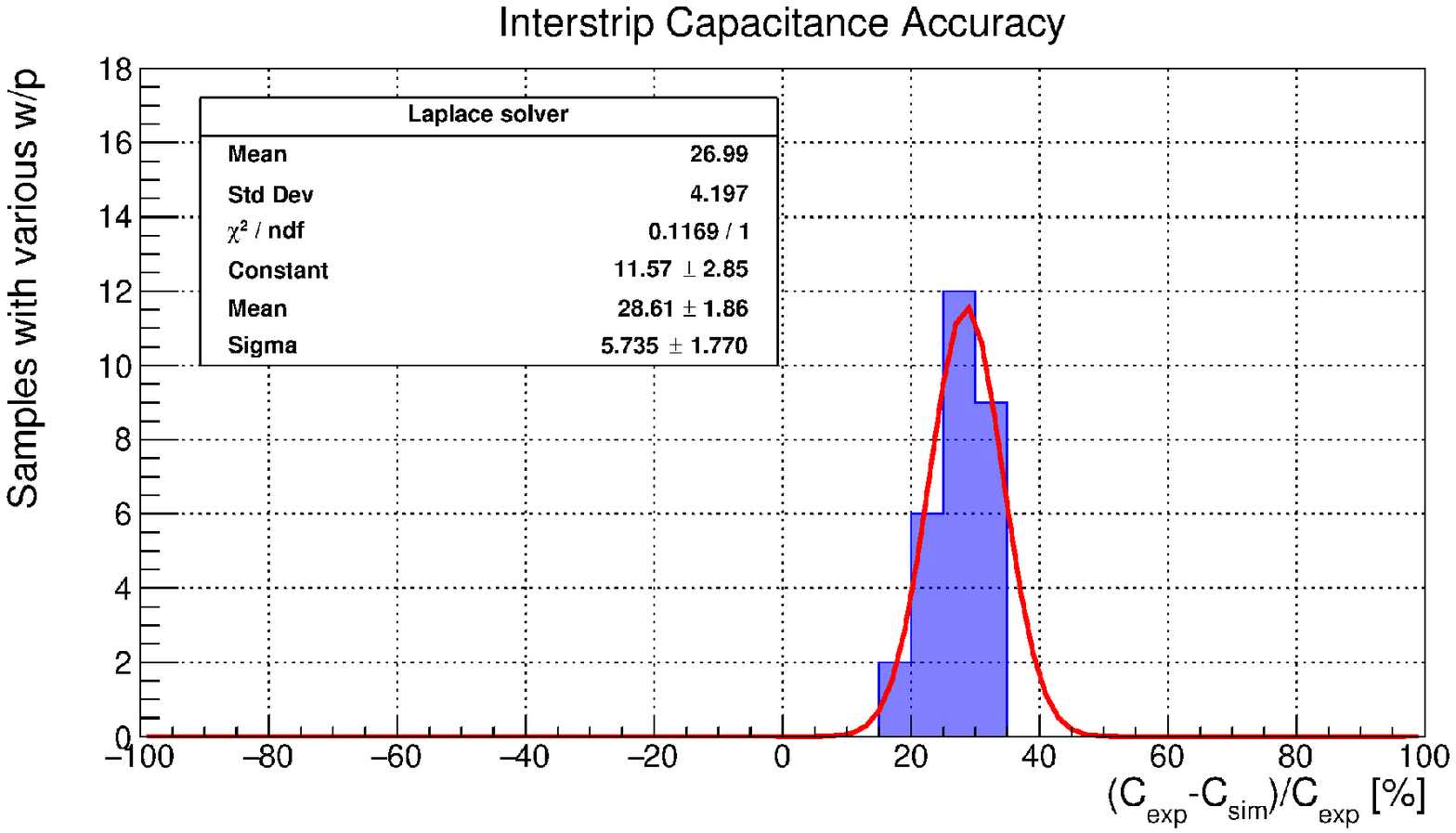}
		\caption{
			\label{fig:accuracy_histo_Cint_Laplace}}      
	\end{subfigure}\quad
	\begin{subfigure}{0.44\textwidth}
		\centering
		\includegraphics[height=4.0cm,width=\textwidth]{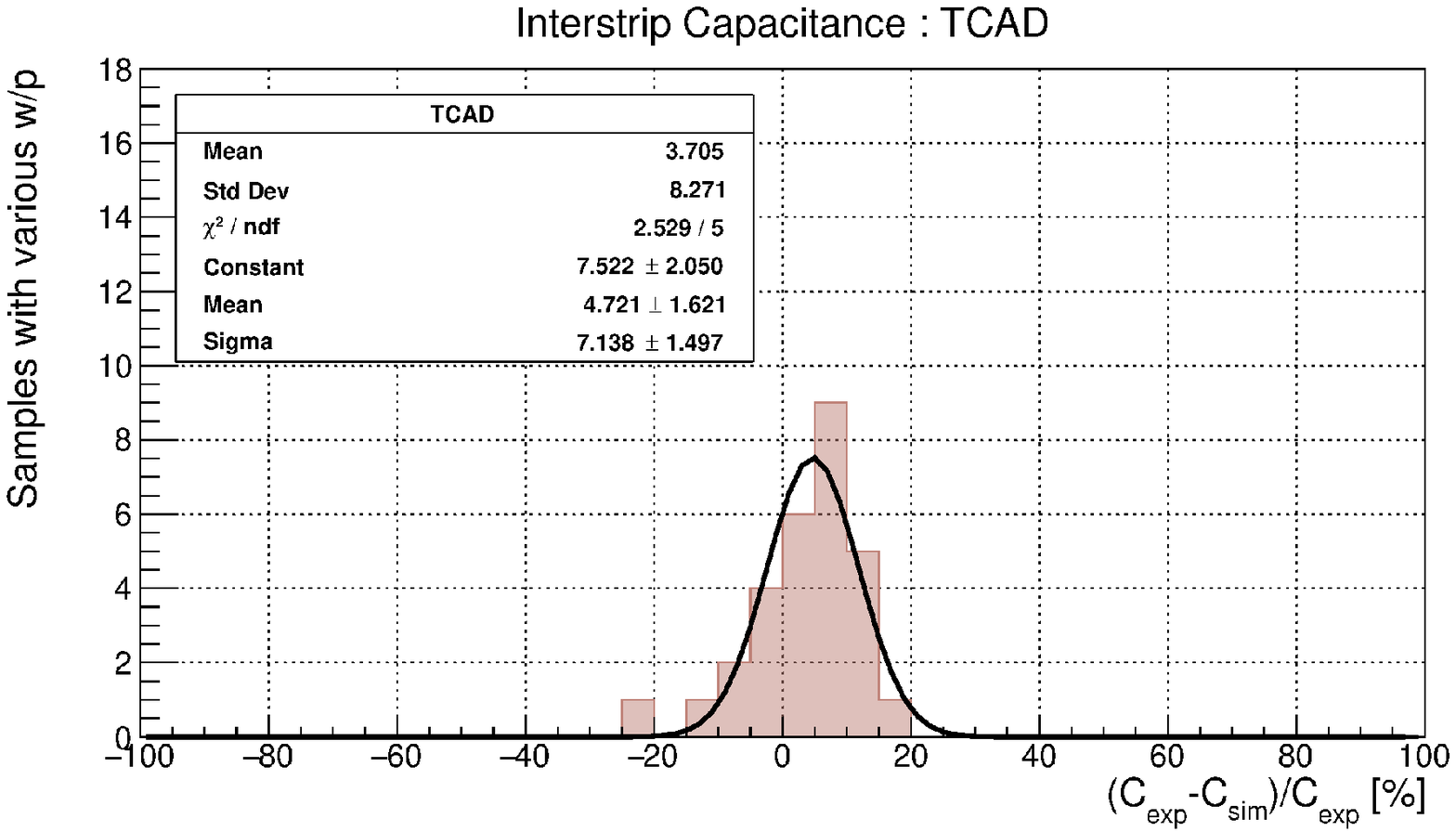}
		\caption{ \label{fig:accuracy_histo_Cint_TCAD}}
	\end{subfigure}
	\caption[]{Histograms of the relative error of the simulated results to the experimental values for the interstrip capacitance for the Laplace solver (figure \ref{fig:accuracy_histo_Cint_Laplace}) and for the TCAD simulations (figure \ref{fig:accuracy_histo_Cint_TCAD}).
	}	
\end{figure}

Figures \ref{fig:accuracy_histo_Cint_Laplace} and \ref{fig:accuracy_histo_Cint_TCAD} show the relative errors of the simulated results ($C_{sim}$) from our Laplace solver and the TCAD simulations respectively, to the experimental values ($C_{exp}$) of the interstrip capacitance as it has been calculated with our Laplace solver and with the TCAD simulations, respectively. The numerical calculations made with the Laplace solver have a mean value of $27\%$ with a standard deviation of $4\%$ while the calculations made with TCAD have mean value of $4\%$ and a standard deviation of $8\%$.

It must be noted that the numerical calculations from the Laplace solver are tailored to the ideal case where the sensor is free from static charges. In addition, no oxide and no aluminum contacts above the strip plane are taken into account. Thus, it calculates only the capacitance between adjacent implants $C_{I_{i}-I_{j}}$. On the other hand, TCAD makes a more detailed simulation with more accurate physical models. Moreover, the interstrip capacitances are calculated by following equation \ref{eq:Cint} where the capacitances between adjacent aluminum $C_{M_{i}-M_{j}}$ and  between adjacent aluminum and implants  $C_{M_{i}-I_{j}}$, $C_{I_{i}-M_{j}}$ are taken into account. In ref. \cite{CHATTERJI20031491} it is noted that when the metal overhang is absent the implant-implant capacitance is the dominant component of the total interstrip capacitance. Otherwise, when the overhang is present and begins to increase, the other three components of the interstrip capacitance start to increase with a simultaneous decrease of the implant-implant capacitance. 

\begin{figure}[htbp!]
		\centering
				\includegraphics[height=4.0cm  , width=0.5\textwidth]{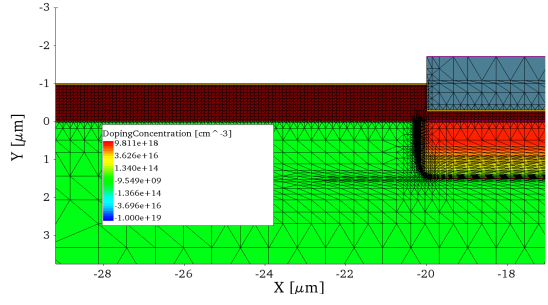}
	\caption[]{The simulated structure without metal-overhang close to one strip edge. The n-type implant is
		displayed in red which is simulated with a Gaussian profile, the aluminum contacts are displayed
		with gray, the SiO\textsubscript{2} is displayed with brown and the p-type silicon bulk is displayed with green.}
\label{fig:MSSD_FZ120_CloseStrip_without_overhang}
\end{figure}

The numerical calculations as described in section \ref{sbsc:numerical_solution_strip_sensors} should calculate the interstrip capacitances in the case where the metal overhang is absent and thus the implant-implant component of the interstrip capacitance is dominant. In order to check the validity of our software in calculating the implant-implant component of the capacitance we have simulated also the case where the metal overhang is absent with TCAD. In this case the aluminum width is taken to be equal to implant width (figure \ref{fig:MSSD_FZ120_CloseStrip_without_overhang}). The other properties of the simulated structure were kept the same as described in section \ref{sbsc:TCAD_simulations_strip_sensors}. 

\begin{figure}[!htb]
   	\centering
   	\begin{subfigure}{0.31\textwidth}
   		\centering
   		\includegraphics[height=3.5cm,width=\textwidth]{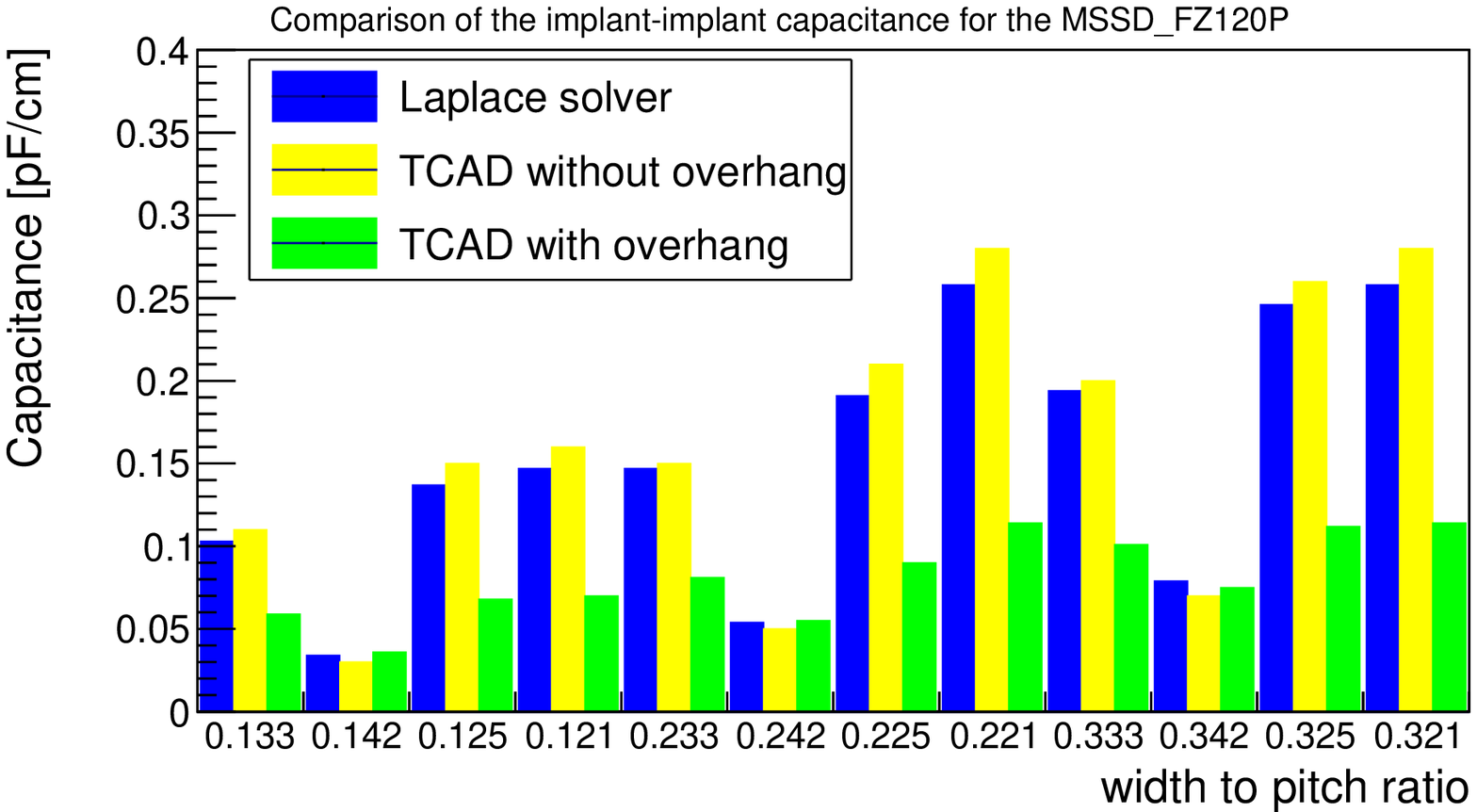}
   		\caption{
   		\label{fig:Comparison_Cimim_MSSD_120P}}      
   	\end{subfigure}
   	\quad
   	\begin{subfigure}{0.31\textwidth}
   		\centering
   		\includegraphics[height=3.5cm, width=\textwidth]{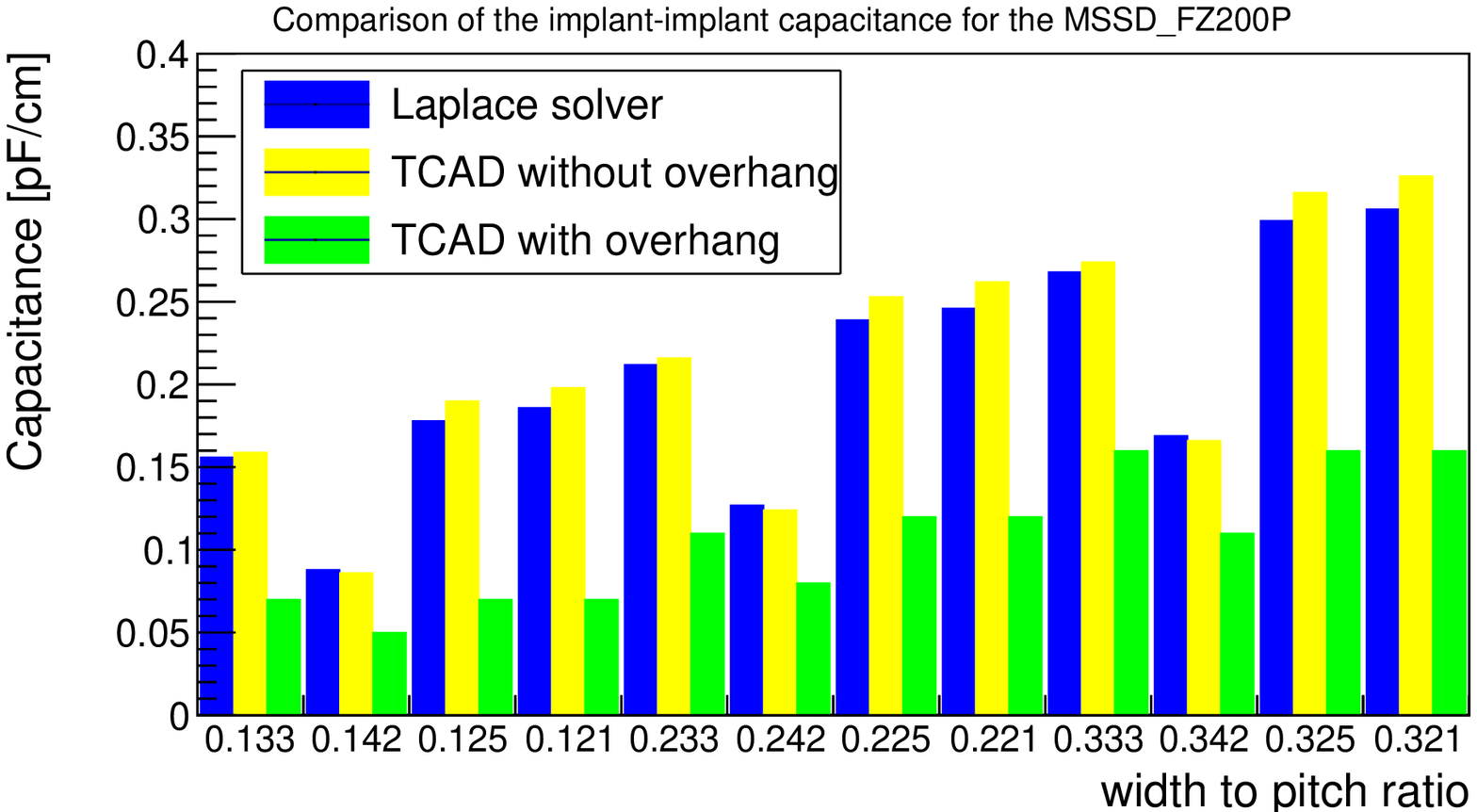}
   		\caption{ \label{fig:Comparison_Cimim_MSSD_200P}}
   	\end{subfigure}\quad
   	\begin{subfigure}{0.31\textwidth}
   	\centering
   	\includegraphics[height=3.5cm  , width=\textwidth]{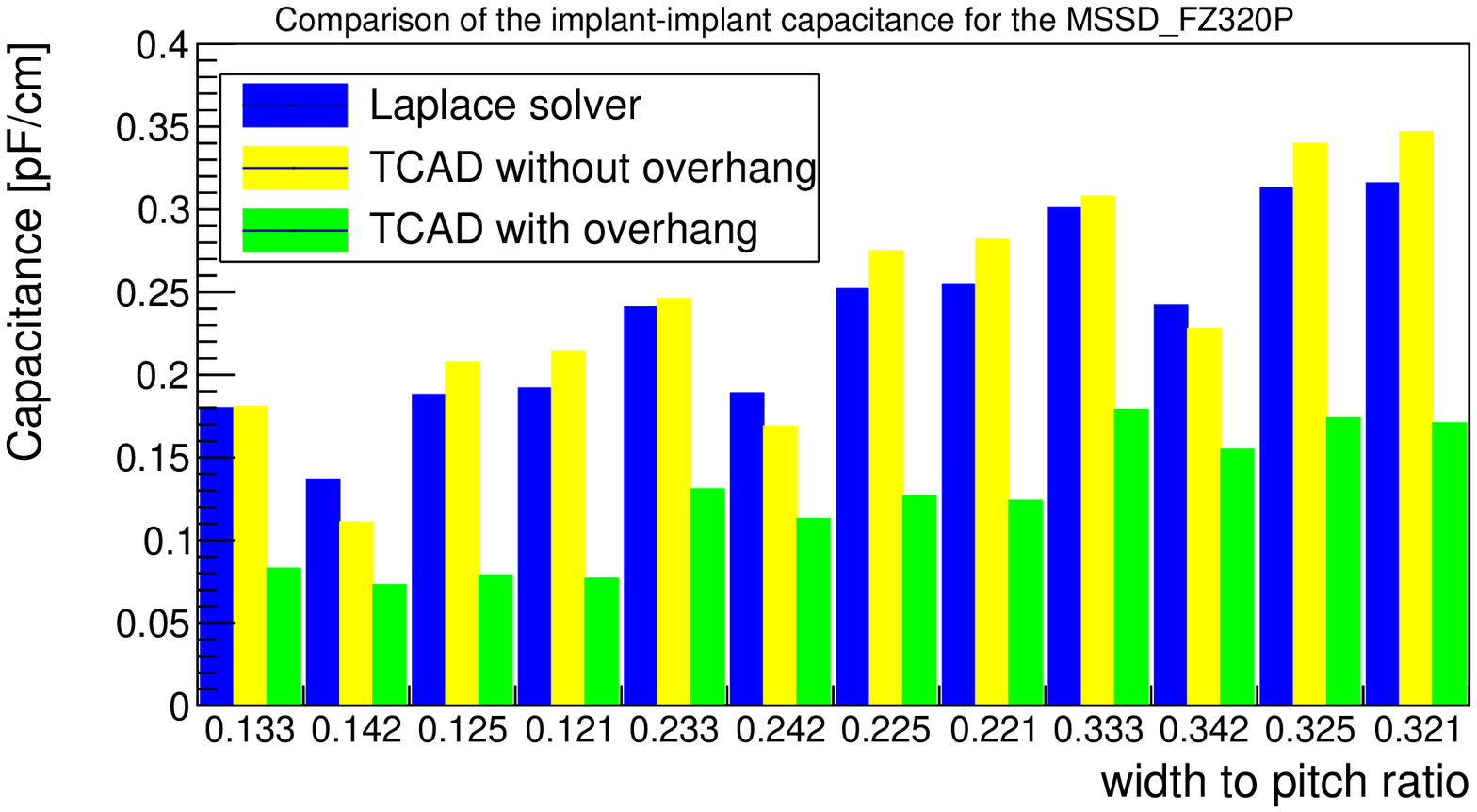}
   	\caption{\label{fig:Comparison_Cimim_MSSD_320P} }  
   \end{subfigure}
	\caption[]{Comparison of the calculted results (blue) with TCAD simulations in the case were metal-overhang is present (green) and without metal-overhang (yellow) for the FZ120P \ref{fig:Comparison_Cimim_MSSD_120P}, FZ200P \ref{fig:Comparison_Cimim_MSSD_200P} and FZ320P
		\ref{fig:Comparison_Cimim_MSSD_320P} sensors.}	
   \end{figure}

Histograms,
\ref{fig:Comparison_Cimim_MSSD_120P}, \ref{fig:Comparison_Cimim_MSSD_200P},
\ref{fig:Comparison_Cimim_MSSD_320P} show the simulated results of the implant-implant component of the  capacitances for two cases with metal-overhang (depicted in green) and without metal-overhang (depicted in yellow), compared with results from Laplace solver(depicted in blue).

 The calculated results of our software agree with the simulated results for the implant-implant components of the interstrip capacitance in the case were  metal overhang is absent with an average accuracy of  $\approx 7$ $\%$. The implant-implant component of the capacitance decreases in the case where metal overhang is present as it was expected according to \cite{CHATTERJI20031491}. Also changing the values of the dielectric constant of the ambient space to that of $SiO_{2}$ doesn't improve the results. However, it sightly decreases the interstrip capacitance while the backplane remains the same.  

\subsection{Comparison of 3D Laplace solver with experimental data of various pixel geometries}

\begin{figure}[!htbp]	
		 \includegraphics[height=5.0cm  , width=0.6\textwidth]{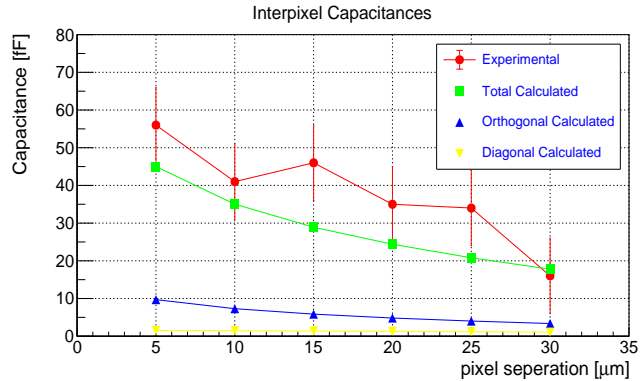}	
		\caption[]{Total interpixel capacitance for sensors with fixed pitch=100 \si{\micro\meter} and varying separation gap in comparison with experimental data extracted from  \cite{Bonvicini}. Also the calculated results for the orthogonal and the diagonal capacitances are shown.} 
	\label{fig:Ctip_pitch100}
\end{figure}

In ref. \cite{Bonvicini}, pixel capacitances have been measured on pixel sensors from 6 different n-type wafers with fixed pitch=100 \si{\micro\meter} and with varying separation gap between the pixel implants from 5 \si{\micro\meter} to 30 \si{\micro\meter} with a 5 \si{\micro\meter} gap step, resulting in an implant width that ranges from 95 \si{\micro\meter} to  70 \si{\micro\meter}. For comparison we have calculated the capacitances for those sensors with our 3D Laplace solver. Figure \ref{fig:Ctip_pitch100} shows the calculated  total interpixel capacitances $C_{tip}$ for the six different structures in comparison with the literature measurements taken from ref. \cite{Bonvicini}. In this case, due to the square geometry of the pixels and with reference to figure \ref{fig:pixel9rectangle}, $C_{01}=C_{02}$, while with reference  to figure 2 of  \cite{Bonvicini}, $C_{01} \equiv C_{ip}$, which is the orthogonal interpixel capacitance and $C_{03} \equiv C_{diag}$ which is the diagonal interpixel capacitance. The data are compared with the total interpixel capacitance, as it is sensed by a virtually grounded preamplifier, given by equation \ref{eq:interpixel_capacitance}. The orthogonal ($C_{01}$ ,$C_{02}$) and diagonal $C_{03}$ components are drawn as well for clarity.
\begin{equation}
\label{eq:interpixel_capacitance}
C_{tip}=2C_{01}+2C_{02}+4C_{03}
\end{equation} 

The calculated results approximate the lower error limit of the  experimental data of the total interpixel capacitance. The diagonal capacitance in ref. \cite{Bonvicini} was approximated to $2.7$ \si{\femto\farad} and $3.2$ \si{\femto\farad} on a second fit, which excludes the last experimental data. In our calculations the diagonal capacitance is varied between $1.5$ \si{\femto\farad} for  $5$ \si{\micro\meter} gap  and $1.0$ \si{\femto\farad} for  $30$ \si{\micro\meter} gap. The orthogonal interpixel capacitance in \cite{Bonvicini} range, between $1-2$ \si{\femto\farad} (30 \si{\micro\meter} gap) to $11-12$ \si{\femto\farad} (5 \si{\micro\meter} gap) while from our calculations the orthogonal capacitance range between $3.37$ \si{\femto\farad} ($30$ \si{\micro\meter} gap) to $9.70$ \si{\femto\farad} ($5$ \si{\micro\meter} gap).

\begin{figure}[!htbp]	
		 \includegraphics[height=4.5cm  , width=0.6\textwidth]{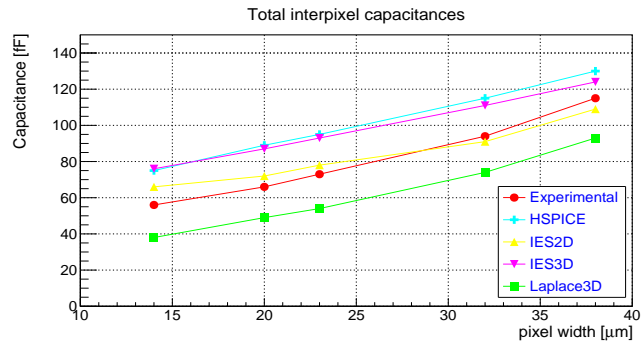}	
		\caption[]{Experimental and simulated data with HSPICE ,IES Coulomb in two and three dimensions of LBNL n-type unirradiated sensors extracted from table 2 and table 8 of ref. \cite{GORFINE2001336} in comparison with results from Laplace solver in  three dimensions for the total interpixel capacitance.} 
	\label{fig:total_interpixel_LBNL}
\end{figure}

In ref. \cite{GORFINE2001336} pixel capacitances have been measured for irradiated and non-irradiated sensors from LBNL and Atlas/LHC test structures along with simulations with HSPICE \cite{HSPICE} and IES Coulomb \cite{IESCoulomb} in two and three dimensions. Each structure includes six $3\times9$ arrays of rectangular pixels with a $50$ \si{\micro\meter} pitch in their short direction and $536$ \si{\micro\meter} pitch in their long direction. The implant width and the separation gap vary in each test structure. We have calculated the capacitances for the LBNL n-type test structures  with  the 3D Laplace solver. The experimental and simulated data for the total interpixel capacitance (equation \ref{eq:interpixel_capacitance}) reported in \cite{GORFINE2001336} in comparison  with calculations made with the 3D Laplace solver are presented in figure \ref{fig:total_interpixel_LBNL}. The Laplace solver results agree with the measurements with a relative error which is less than $32 \%$.

\begin{figure}[!htbp]	
		 \includegraphics[height=4.5cm  , width=0.6\textwidth]{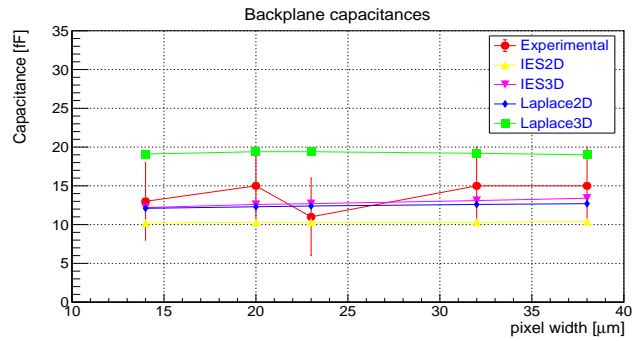}	
		\caption[]{Experimental and simulated data with IES Coulomb in two and three dimensions of LBNL n-type unirradiated sensors extracted from table 6 and of ref. \cite{GORFINE2001336} in comparison with results from Laplace solver in two and three dimensions for the backplane capacitance.} 
	\label{fig:Backplane_LBNL}
\end{figure}

Figure \ref{fig:Backplane_LBNL} shows the experimental results for the backplane capacitance along with simulation results from IES Coulomb in two and three dimensions and with calculations with our software by using both the two and three dimensional methods. The calculated results from our software are inside the error assessment of $\pm 5$ \si{\femto\farad}  that is noted in ref. \citep{GORFINE2001336} expect for arrays 4 and 6 in the 3D case. 

\begin{figure}[!htbp]
	\centering
	\begin{subfigure}{0.48\textwidth}
		\centering
	 \includegraphics[height=4.5cm  , width=\textwidth]{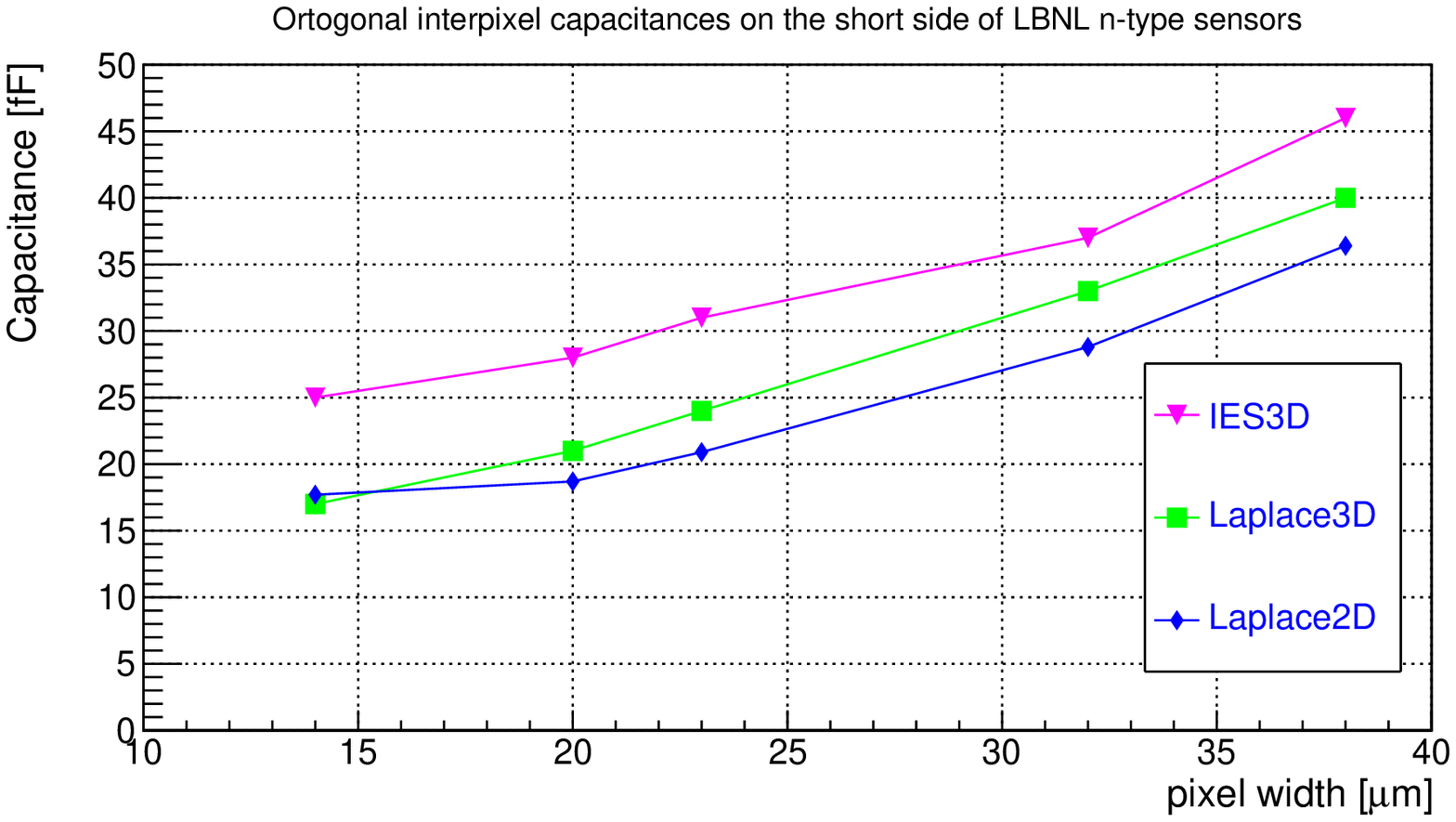}	
		\caption[]{} 
	\label{fig:Orthogonal_LBNL}
	\end{subfigure}
	\quad
	\begin{subfigure}{0.48\textwidth}
		\centering
		 \includegraphics[height=5.0cm  , width=\textwidth]{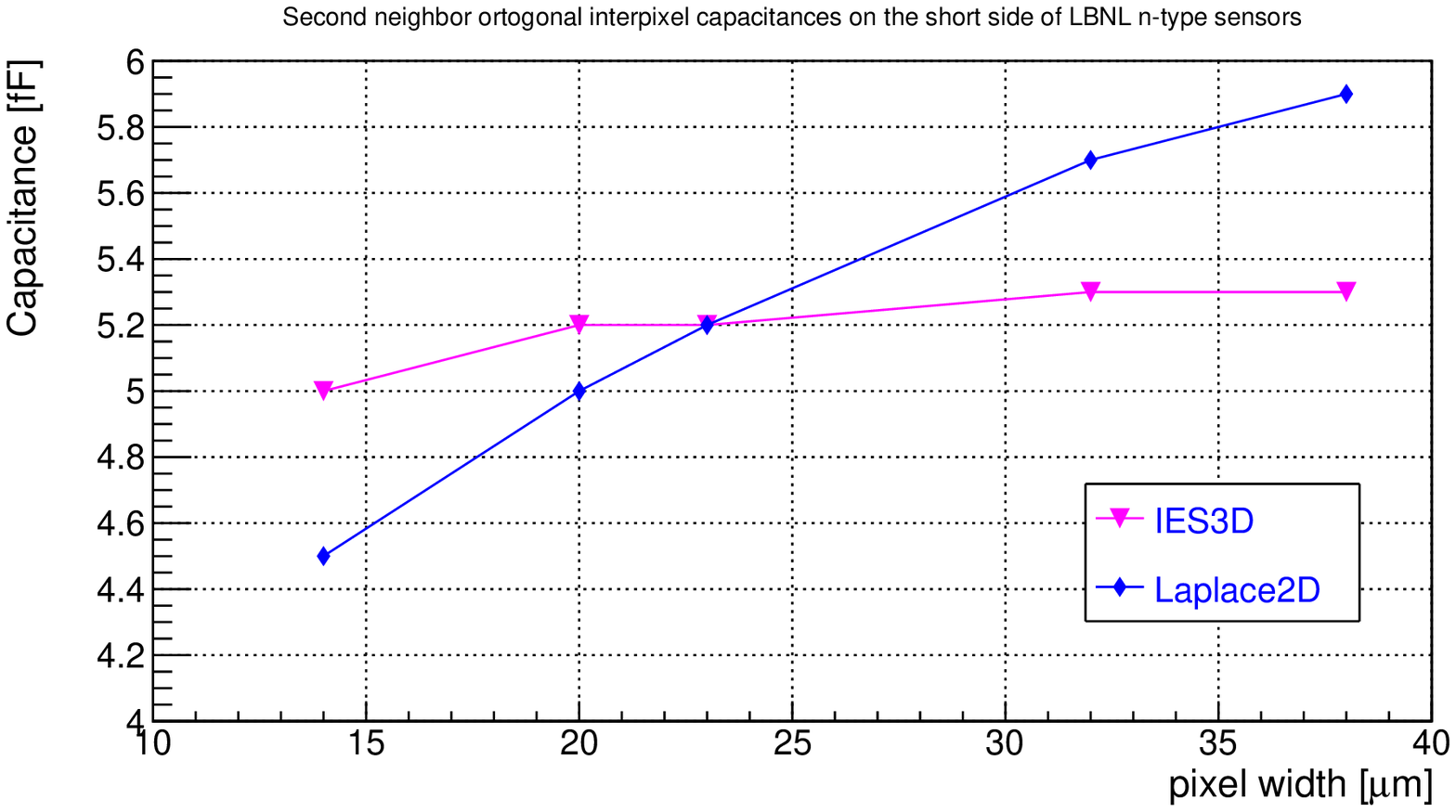}	
		\caption[]{} 
	\label{fig:Second_Orthogonal_LBNL}
	\end{subfigure}	
		\caption[]{Comparison of simulated data for the orthogonal interpixel capacitance on the short side with results from Laplace solver in three and two dimensions (figure \ref{fig:Orthogonal_LBNL}) and comparison of simulated data for the second neighbor capacitance on the short side of LBNL n-type sensors (figure \ref{fig:Second_Orthogonal_LBNL}). The simulated data were extracted from table 9 of ref. \cite{GORFINE2001336}. 
		}
	\label{fig:orthogonal_and_second_neighbor_LBNL}	
\end{figure}

The orthogonal interpixel capacitances on the short side are compared with simulated results from IES Coulomb. These results are presented in figure \ref{fig:Orthogonal_LBNL}. Also in ref. \cite{GORFINE2001336} the second neighbor interpixel capacitances have been calculated. By using the 2D Laplace solver these capacitances were calculated. These results are shown in figure \ref{fig:Second_Orthogonal_LBNL}.

\subsection{Comparison between TCAD and 3D Laplace solver for various pixel geometries}

\begin{figure}[!htbp]
	\centering
	\begin{subfigure}{0.48\textwidth}
		\centering
		\includegraphics[height=4.0cm,width=\textwidth]{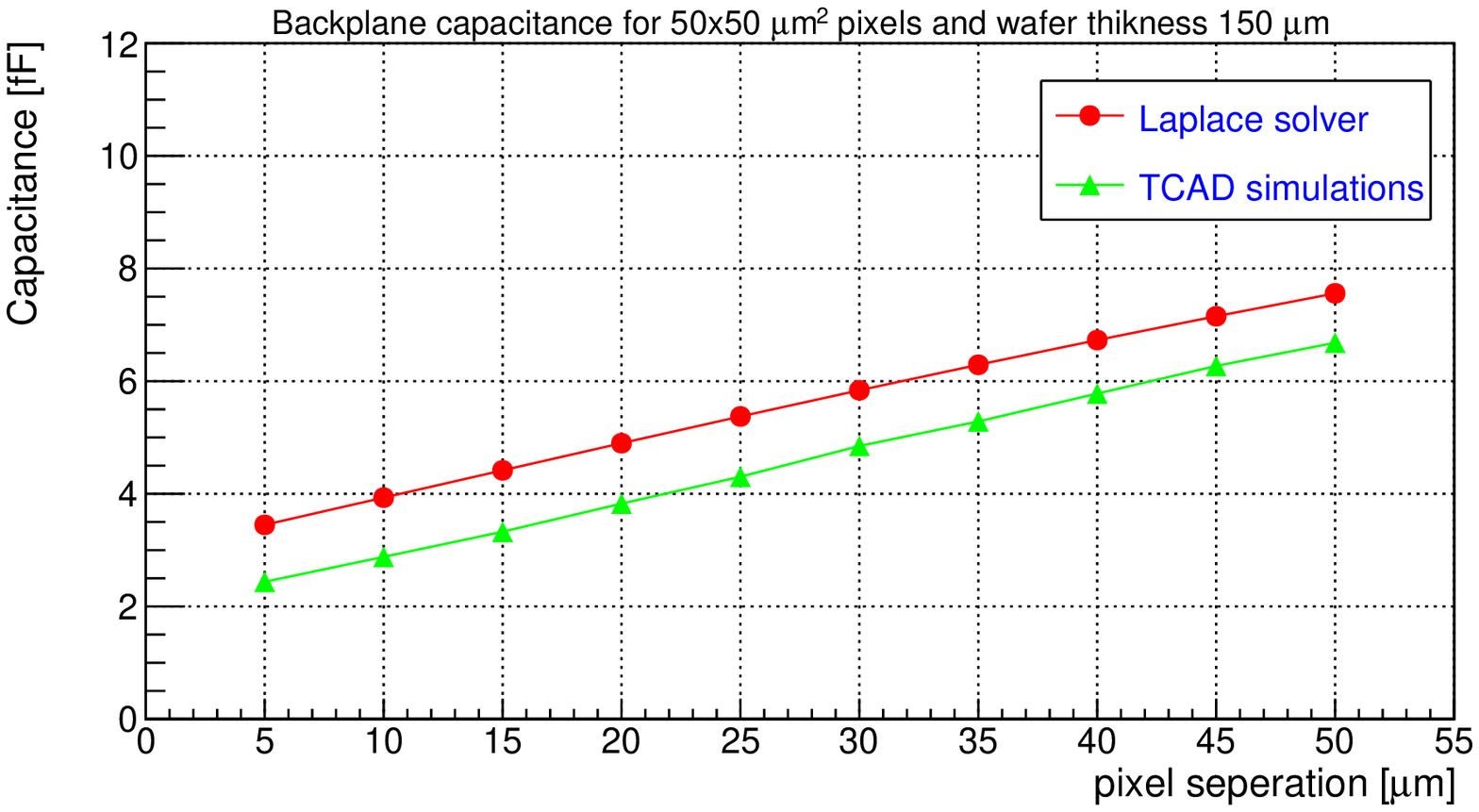}
		\caption{
			\label{fig:Comparison_Cback_50:50}}      
	\end{subfigure}
	\quad
	\begin{subfigure}{0.48\textwidth}
		\centering
		\includegraphics[height=4.0cm  , width=\textwidth]{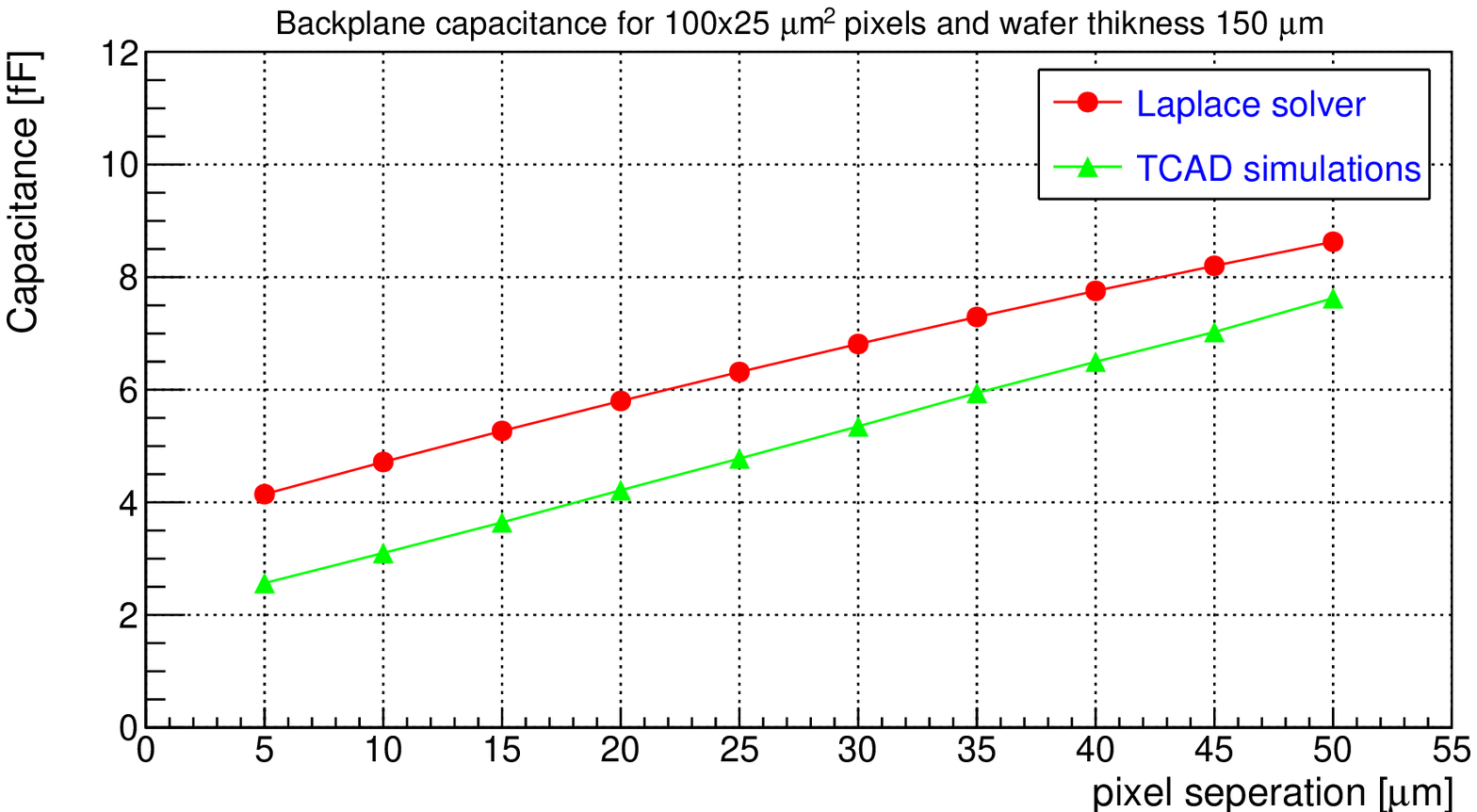}
		\caption{ \label{fig:Comparison_Cback_100:25}}
	\end{subfigure}	
		\caption[]{ Simulated results by using TCAD (green line) for the backplane  capacitance compared with  simulated   results from our program (red line) which implements the numerical method described in \ref{sbsc:numerical_solution_pixel_sensors} for sensors with $50\times50$ \si{\micro\meter^2} \ref{fig:Comparison_Cback_50:50} and $100\times25$ \si{\micro\meter^2} pixel area respectively \ref{fig:Comparison_Cback_100:25}.}
	\label{fig:Comparison_Cback}	
\end{figure}

We have simulated the capacitances for pixel sensors with pixel geometries $50\times50$ \si{\micro\meter^2} and $100\times25$ \si{\micro\meter^2} and thicknesses of 150 \si{\micro\meter}. These configurations are appropriate to the developmental work in progress for the Phase-2 upgrade of the pixel systems in the CMS/LHC and the ATLAS/LHC experiment at CERN. The separation gaps vary between $5$ \si{\micro\meter} to $50$ \si{\micro\meter} with a $5$ \si{\micro\meter} step size. Simulations were performed by using both TCAD and the 3D Laplace solver. Figure \ref{fig:Comparison_Cback} shows the simulated results for the backplane capacitances for sensors with $50\times50$ \si{\micro\meter^{2}} pixels and for sensors with $100\times25$ \si{\micro\meter^{2}} pixels. The backplane capacitance calculated with the Laplace solver is systematically larger than the one obtained from the TCAD simulations in all the cases for about $1.5$ \si{\femto\farad}. A possible reason for this is the better description of the deep diffusion on the backplane that is simulated with TCAD. 

\begin{figure}[!htbp]
	\centering
	\begin{subfigure}{0.48\textwidth}
		\centering
		\includegraphics[height=4.0cm,width=\textwidth]{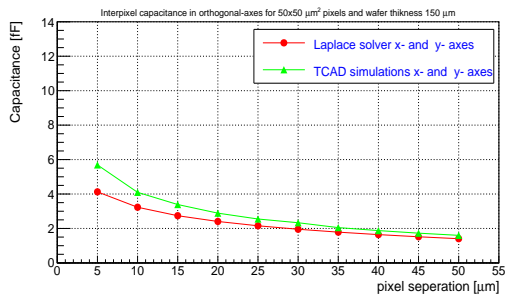}
		\caption{
			\label{fig:Comparison_Cint_50:50}}      
	\end{subfigure}
	\quad
	\begin{subfigure}{0.48\textwidth}
		\centering
		\includegraphics[height=4.0cm  , width=\textwidth]{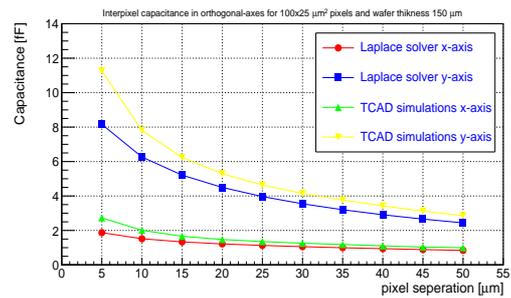}
		\caption{ \label{fig:Comparison_Cint_100x25}}
	\end{subfigure}	
		\caption[]{Simulated results by using TCAD (green-yellow lines) for the orthogonal interpixel  capacitance in x- and y- axes, compared with  simulated   results from our program (red-blue lines) for sensors with $50\times50$ \si{\micro\meter^2} \ref{fig:Comparison_Cint_50:50} and $100\times25$ \si{\micro\meter^2} pixel area respectively \ref{fig:Comparison_Cint_100x25} .}
	\label{fig:Comparison_Cint}	
\end{figure}

\begin{figure}[!htbp]
	\begin{subfigure}{0.48\textwidth}
		\centering
		\includegraphics[height=4.0cm,width=\textwidth]{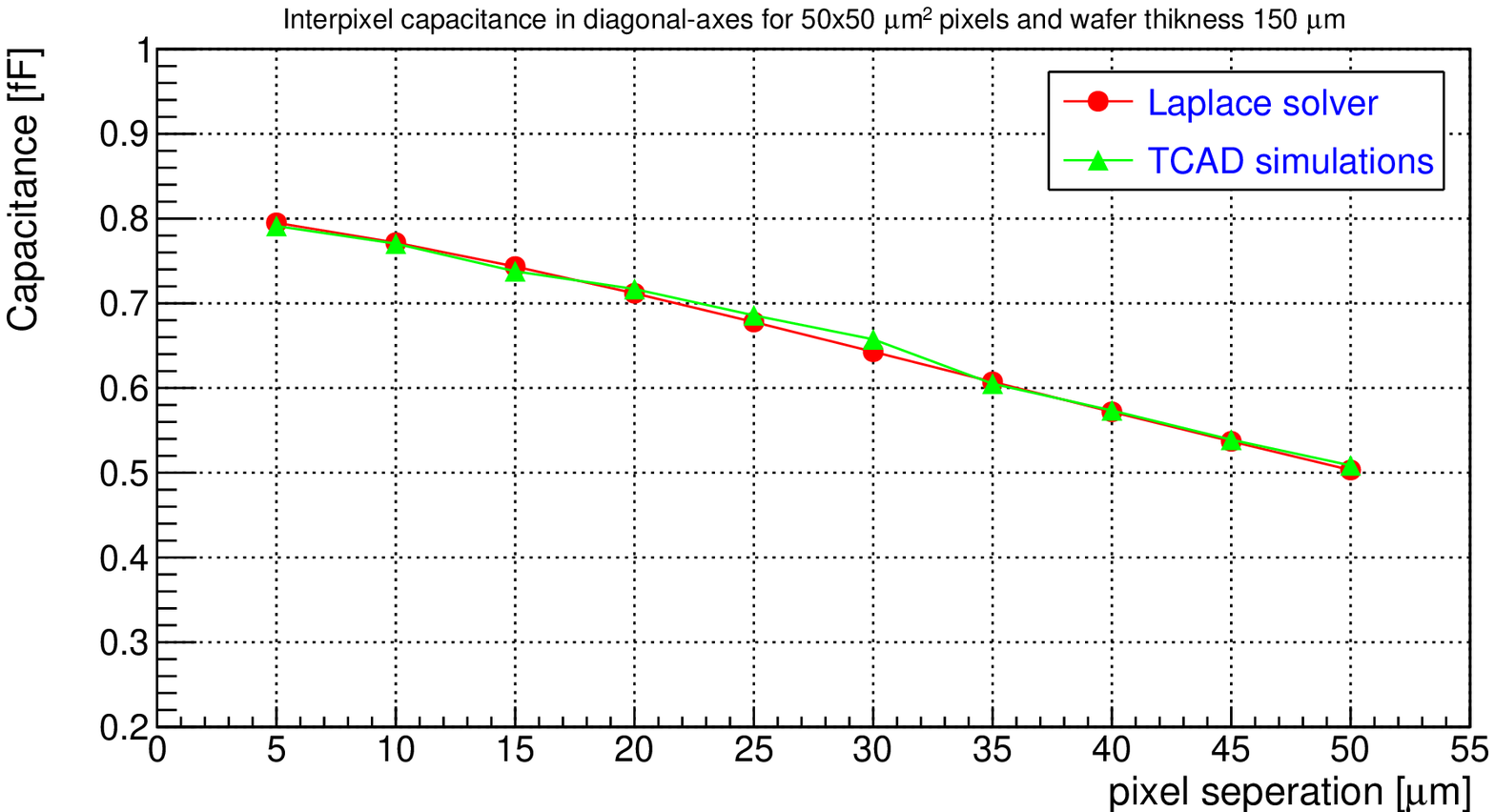}
		\caption{
			\label{fig:Comparison_Cdiag_50:50}}      
	\end{subfigure}
	\quad
	\begin{subfigure}{0.48\textwidth}
		\centering
		\includegraphics[height=4.0cm  , width=\textwidth]{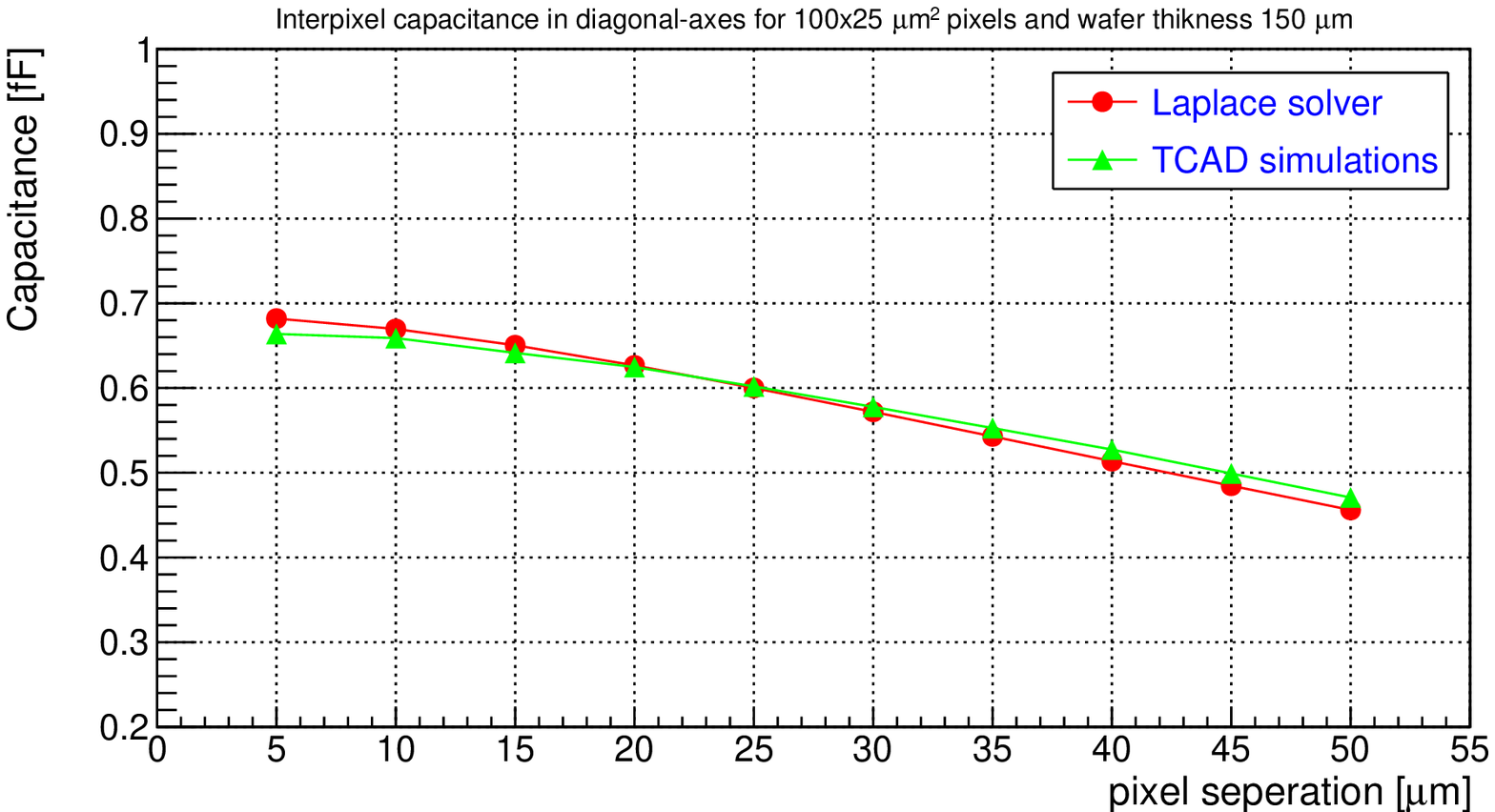}
		\caption{ \label{fig:Comparison_Cdiag_100x25}}
	\end{subfigure}	
		\caption[]{Simulated results by using TCAD (green line) for the diagonal interpixel  capacitance, compared with  simulated   results from our program (red line)  for sensors compared with  simulated   results from our program (red-blue lines) for sensors with $50\times50$ \si{\micro\meter^2} \ref{fig:Comparison_Cdiag_50:50} and $100\times25$ \si{\micro\meter^2} pixel area respectively \ref{fig:Comparison_Cdiag_100x25} .}
	\label{fig:Comparison_Cdiag}	
\end{figure}

\begin{figure}[!htbp]
	\centering
	\begin{subfigure}{0.48\textwidth}
		\centering
		\includegraphics[height=4.0cm,width=\textwidth]{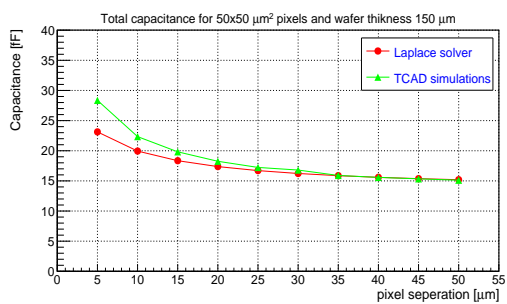}
		\caption{
			\label{fig:Comparison_Ctot_50:50}}      
	\end{subfigure}
	\quad
	\begin{subfigure}{0.48\textwidth}
		\centering
		\includegraphics[height=4.0cm  , width=\textwidth]{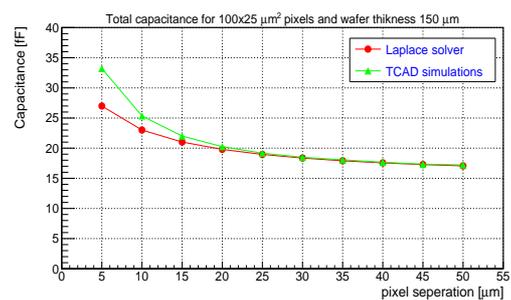}
		\caption{ \label{fig:Comparison_Ctot_100x25}}
	\end{subfigure}	
\caption[]{Simulated results by using TCAD (green line) for the total capacitance, compared with  simulated   results from our program (red line) for sensors compared with  simulated   results from our program (red-blue lines) for sensors with $50\times50$ \si{\micro\meter^2} \ref{fig:Comparison_Ctot_50:50} and $100\times25$ \si{\micro\meter^2} pixel area respectively \ref{fig:Comparison_Ctot_100x25}.}	
	\label{fig:Comparison_Ctot}
\end{figure}

Figures \ref{fig:Comparison_Cint} and \ref{fig:Comparison_Cdiag} show the simulated results for the interpixel capacitances. Both simulations show
a very good agreement in the calculation of the orthogonal and diagonal interpixel capacitances especially for larger separation gaps. Figure \ref{fig:Comparison_Ctot} shows the simulated results for the total capacitances, where the total capacitance is the sum of all the total interpixel capacitance (equation \ref{eq:interpixel_capacitance}) including the backplane. Again the two simulations agree well in the calculation of the total capacitance especially for gaps larger than $15$  \si{\micro\meter}.

\begin{figure}[!htbp]
	\centering
	\begin{subfigure}{0.48\textwidth}
		\centering
		\includegraphics[height=4.5cm,width=\textwidth]{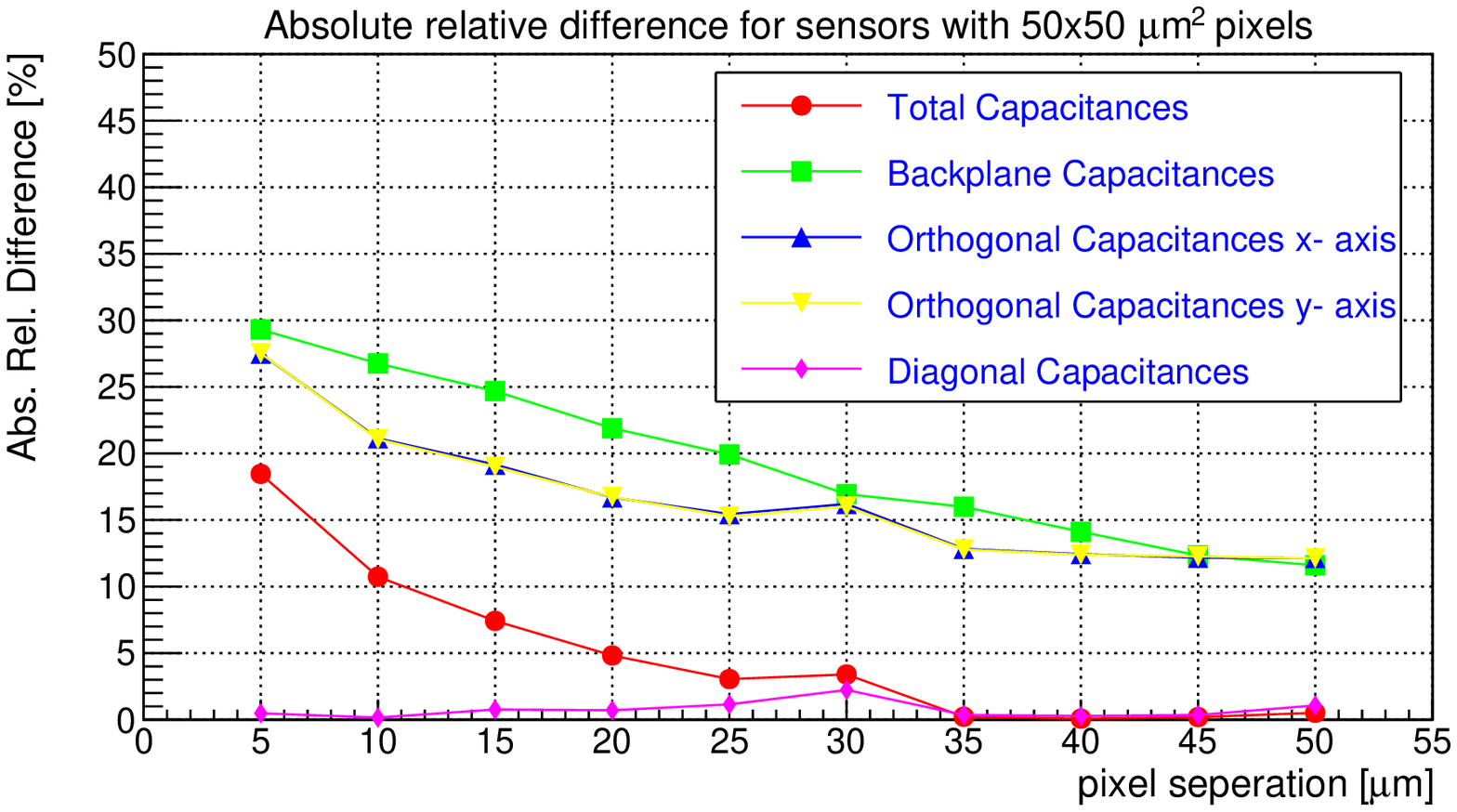}
		\caption{
			\label{fig:Abs.Rel.Diff_50x50}}      
	\end{subfigure}
	\quad
	\begin{subfigure}{0.48\textwidth}
		\centering
		\includegraphics[height=4.5cm  , width=\textwidth]{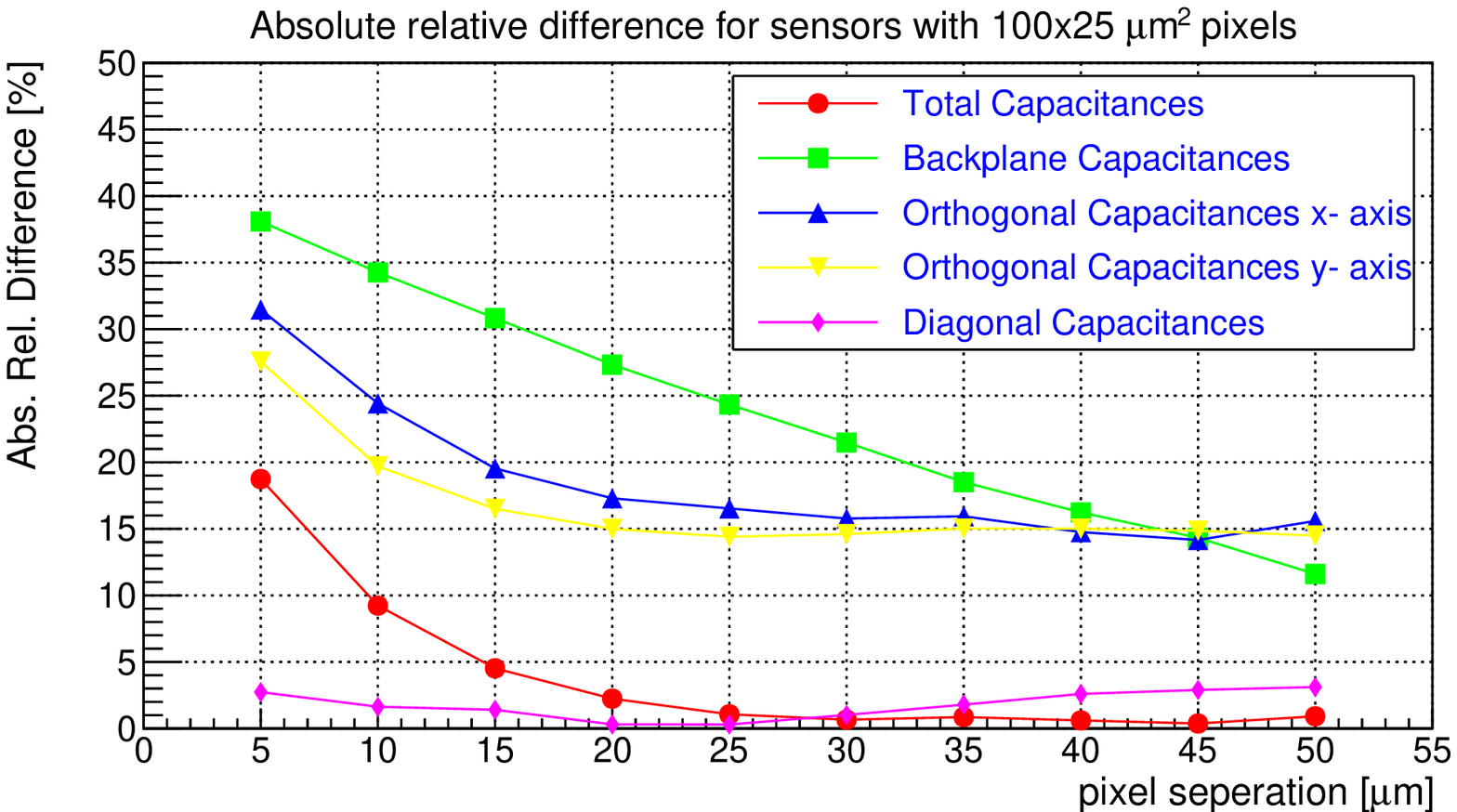}
		\caption{ \label{fig:Abs.Rel.Diff_100x25}}
	\end{subfigure}	
		\caption[]{ Absolute relative difference between the calculated results with Laplace solver and the TCAD simulation results for sensors with $50\times50$ \si{\micro\meter^2} \ref{fig:Abs.Rel.Diff_50x50} and $100\times25$ \si{\micro\meter^2} pixel area respectively \ref{fig:Abs.Rel.Diff_100x25}. 
		}
	\label{fig:fig:Abs.Rel.Diff}	
\end{figure}

Figures \ref{fig:Abs.Rel.Diff_50x50} and \ref{fig:Abs.Rel.Diff_100x25} show the relative difference between the calculated results with Laplace solver and TCAD simulation results, respectively. It can be noted that the difference between the two simulations decrease with the increase of the separation gap. The relative difference for the backplane capacitance  ranges between 29 $\%$ and 12 $\%$ for sensors with $50\times50$ \si{\micro\meter^2} pixels and between 38 $\%$ and 12 $\%$ for
sensors with $100\times25$ \si{\micro\meter^2} pixels. For the orthogonal interpixel capacitances it ranges between 28 $\%$ ans 12 $\%$ for sensors with $50\times50$ \si{\micro\meter^2} pixels, while for the diagonal capacitances it is less than 2 $\%$ and for
sensors with $100\times25$ \si{\micro\meter^2} pixels it ranges between 32 $\%$  and 16 $\%$  for x- axis and between 28 $\%$ and 15 $\%$ for y- axis, while for the diagonal capacitances it is less than 3 $\%$. For total capacitance it ranges between 19 $\%$ and 0.5 $\%$ for sensors with $50\times50$ \si{\micro\meter^2} pixels and between 19 $\%$ and 0.9 $\%$ for
sensors with $100\times25$ \si{\micro\meter^2} pixels.

\subsection{Comparison of the runtime between the two simulations}
Table \ref{tab:time_vs_resolution} shows the time that is needed to calculate the capacitances by using the Laplace solver. The variable resolution is the number of elements that are used for the numerical calculations. Two examples are given for comparison with TCAD running on an 8-core processor at 3.70 $GHz$. In the case of strip sensors, for a 2D Laplace resolution of 3.5 $10^{5}$, the run time is 7.3 \si{\second}, while for the TCAD simulation with the same resolution the runtime is 30 \si{\minute}. In the pixel case, for a 3D Laplace resolution of 2.0 $10^{6}$ the run time is 34.9 \si{\second}, while for the TCAD simulations with the same resolution the runtime is 4 \si{\hour}.

\begin{table}[!htbp]
	\centering
		\caption[]{Calculation time for different values of discretization for pixel and strip sensors.}
		\begin{tabular}{r|l|ll}
			\hline 
			& resolution& time \si{\second}  \\
			\hline   \hline
			&2.6 $10^{5}$&   4.9 \\
			pixel  & 1.1 $10^{6}$&19.2\\ 
			 & 2.0 $10^{6}$& 34.9\\ 
			&4.2 $10^{6}$&   74.1 \\
			\hline 
			&1.0 $10^{3}$ &  0.2\\
			strip &1.6 $10^{4}$& 1.7\\
			&6.6 $10^{4}$& 5.3\\
			&3.5 $10^{5}$& 7.3\\
			&5.2 $10^{5}$&14.2\\
			\hline
		\end{tabular}
	\label{tab:time_vs_resolution}
\end{table}

\section{Conclusion}

The fast numerical solution of the Laplace's equation  described in this work, gives an accurate approximation of the experimental results and of the TCAD simulations. For the strip sensors the mean value of the relative error of all the regions for the 3 MSSD sensors is $14 \%$ for the backplane capacitance and $27 \%$ for the interstrip capacitance, while the implant-implant component for sensors without overhang is approximated with a mean relative error of $7 \%$ , compared with TCAD  simulation results. For pixel sensors the relative error of the calculations was found to be less than $32 \%$ compared to experimental results found in literature. In addition, compared  to the TCAD simulations the calculated results show a very good agreement, especially the calculations of the interpixel and total capacitances for large inter-pixel gaps. 

As a general conclusion, the program that implements the three and two dimensional numerical solution of the Laplace's equation that is described in this paper can be used in order to provide a fast approximation of detector capacitances for planar silicon strip and pixel sensors before a more detailed simulation with EDA tools is performed. This tool is foreseen to be implemented into a web-based application.   

\appendix
	\section{ Properties used for the TCAD simulations} \label{Appendix A}

\begin{table}[!htbp]
\begin{small}
\begin{center}
\begin{tabular}{ |c|c c c| } 
 \hline
Material & FZ 120P &  FZ 200P &  FZ 320P \\ 
 \hline
 Bulk doping concentration$[cm^{-3}]$&  & $3.5e^{12}$ & \\ 
 Strip doping concentration $[cm^{-3}]$& & $1.0e^{19}$ & \\
 Backplane doping concentration $[cm^{-3}]$& & $1.0e^{19}$ &\\
  p-stop doping concentration $[cm^{-3}]$& & $1.0e^{16}$ & \\
  \hline
 $SiO_{2}$ thickness between strips [\si{\micro\meter}] & & $0.95$ & \\
 $SiO_{2}$  thickness between metal-strip [\si{\micro\meter}]  & & $0.25$ & \\
 $SiN_{4}$ thickness  [\si{\micro\meter}]  & & $0.05$ & \\
  Aluminum thickness [\si{\micro\meter}]  & & $0.7$ & \\
  Strip implant thickness [\si{\micro\meter}]  & &$1.5$  & \\
   Error profile backplane depth [\si{\micro\meter}]  & 215 & 125 & 33 \\ 
 \hline
\end{tabular}
\caption{Geometrical properties and doping concentrations used for the TCAD simulation of strip sensors.}
\label{tab:properties_strip_TCAD}
\end{center}
\end{small}
\end{table}

\begin{table}[!htbp]
\begin{small}
		\begin{center}
		\begin{tabular}{ |c| c| } 
			\hline
			Material & n\textsuperscript{+}p \\ 
			\hline
			Bulk doping concentration$[cm^{-3}]$  & $4.0e^{12}$  \\ 
			Pixel doping concentration $[cm^{-3}]$ & $2.0e^{19}$  \\
			Backplane doping concentration $[cm^{-3}]$ & $2.0e^{19}$ \\
			Guard ring doping concentration $[cm^{-3}]$ & $2.0e^{16}$  \\
			\hline
			$SiO_{2}$ thickness  [\si{\micro\meter}]  & $1.0$  \\
			Aluminum thickness [\si{\micro\meter}]   & $0.7$ \\
			Pixel implant thickness [\si{\micro\meter}]   & $1.5$   \\
			Error profile backplane depth [\si{\micro\meter}] & 20 \\ 
			\hline
		\end{tabular}
		\caption{Geometrical properties and doping concentrations used for the TCAD simulation of pixel sensors.}
		\label{tab:properties_pixel_TCAD}
	\end{center}
\end{small}
\end{table}

	\acknowledgments
	  The authors of this paper would like to thank the outer tracker sensor working group of CMS/LHC for providing the MSSD sensors.

		This research is co-financed by Greece and the European Union (European Social Fund- ESF) through the Operational Programme Human Resources Development, Education and Lifelong Learning 2014 2020 in the context of the project "New generation of sensors and electronics for the upgrade of the CMS/LHC experiment at CERN"- MIS 5047807 
	
	\begin{figure}[!htbp]
   \includegraphics[height=2cm,width=\textwidth]{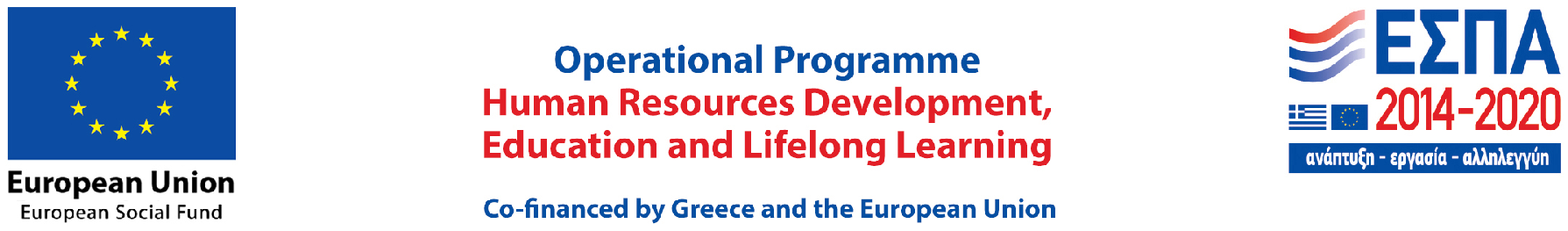}
   \label{fig: 5stripdector}
\end{figure}


	 \bibliographystyle{JHEP}
	 \bibliography{References}

\end{document}